\documentclass[%
preprint,
aps,prb,floatfix,superscriptaddress,showpacs,footinbib,
amsmath,amssymb,
]{revtex4-1}

\usepackage{graphicx}
\usepackage{dcolumn}
\usepackage{bm}
\usepackage[normalem]{ulem}
\usepackage{hyperref}

\usepackage{color}
\newcommand{\bk}{\textbf{k}} \newcommand{\bq}{\textbf{q}}
 
\newcommand{\br}{\textbf{r}}

\newcommand{\mrm}{\mathrm}
\makeatletter
\renewcommand{\fnum@figure}{\textbf{ Figure \thefigure}}
\makeatother

\begin{document}

\title{Extracting phase information about the superconducting order parameter from defect bound states}

\author{Shun Chi}
\affiliation{Department of Physics and Astronomy, University of British Columbia, Vancouver BC, Canada V6T 1Z1}
\affiliation{Stewart Blusson Quantum Matter Institute, University of British Columbia, Vancouver BC, Canada V6T 1Z4}
\author{W. N. Hardy}
\affiliation{Department of Physics and Astronomy, University of British Columbia, Vancouver BC, Canada V6T 1Z1}
\affiliation{Stewart Blusson Quantum Matter Institute, University of British Columbia, Vancouver BC, Canada V6T 1Z4}
\author{Ruixing Liang}
\affiliation{Department of Physics and Astronomy, University of British Columbia, Vancouver BC, Canada V6T 1Z1}
\affiliation{Stewart Blusson Quantum Matter Institute, University of British Columbia, Vancouver BC, Canada V6T 1Z4}
\author{P. Dosanjh}
\affiliation{Department of Physics and Astronomy, University of British Columbia, Vancouver BC, Canada V6T 1Z1}
\affiliation{Stewart Blusson Quantum Matter Institute, University of British Columbia, Vancouver BC, Canada V6T 1Z4}
\author{Peter Wahl}
\affiliation{SUPA, School of Physics and Astronomy, University of St. Andrews, North Haugh, St. Andrews, Fife, KY16 9SS, United Kingdom}
\affiliation{Max-Planck-Institut f\"ur Festk\"orperforschung, Heisenbergstr. 1, D-70569 Stuttgart, Germany}
\author{S. A. Burke}
\affiliation{Department of Physics and Astronomy, University of British Columbia, Vancouver BC, Canada V6T 1Z1}
\affiliation{Stewart Blusson Quantum Matter Institute, University of British Columbia, Vancouver BC, Canada V6T 1Z4}
\affiliation{Department of Chemistry, University of British Columbia, Vancouver BC, Canada V6T 1Z1}
\author{D. A. Bonn}
\affiliation{Department of Physics and Astronomy, University of British Columbia, Vancouver BC, Canada V6T 1Z1}
\affiliation{Stewart Blusson Quantum Matter Institute, University of British Columbia, Vancouver BC, Canada V6T 1Z4}

\date{\today}

\begin{abstract}
Impurity bound states and quasi-particle scattering from these can serve as sensitive probes for identifying the pairing state of a superconducting condensate. We introduce and discuss defect bound state quasi-particle interference (DBS-QPI) imaging as a tool to extract information about the symmetry of the order parameter from spatial maps of the density of states around magnetic and non-magnetic impurities. We show that the phase information contained in the scattering patterns around impurities can provide valuable information beyond what is obtained through conventional QPI imaging. Keeping track of phase, rather than just magnitudes, in the Fourier transforms is achieved through phase-referenced Fourier transforms that preserve both real and imaginary parts of the QPI images. We further compare DBS-QPI to other approaches which have been proposed to use either QPI or defect scattering to distinguish different symmetries of the order parameter.
\end{abstract}

\maketitle

\section{Introduction}
The symmetry of the order parameter in a superconductor is key information required to characterize and understand superconductivity in a material. In conventional superconductors, where the pairing interaction is due to electron-phonon coupling, the pairing symmetry and hence the symmetry of the superconducting order parameter is $s$-wave. For unconventional superconductors, the pairing interaction is believed to be governed by electron-electron interactions, for example, through spin-fluctuation mediated pairing. In this case the strong Coulomb repulsion between the electrons means that non-$s$-wave pairing is favoured, leading to sign-changing order parameters. While for some unconventional superconductors, such as the cuprates or some heavy fermion materials, there is well-established experimental evidence for a specific symmetry of the order parameter, in many cases this is not a settled issue. For the iron-based superconductors, there are only a few compounds where there is consensus that the pairing is of the $s_{\pm}$ type, whereas for others this remains an open question. One reason for uncertainty about the superconducting order parameter is that while many experimental probes are sensitive to the magnitude of the order parameter, there are only very few experimental techniques which can probe the phase and hence completely constrain identification of the symmetry of the order parameter. 

Quasi-particle interference (QPI) imaging, enabled through spectroscopic mapping in a scanning tunneling microscope (STM), has in recent years been established as a powerful tool to characterize electronic states in superconductors as well as a wide range of other materials\cite{hoffman_imaging_2002, wang_quasiparticle_2003, mcelroy_relating_2003, hanaguri_quasiparticle_2007,Rutter2007, kohsaka_how_2008, Roushan2009, Hanaguri2009Dwave, Hanaguri2010,Hess-STM, Allan2012, Allan2013CeCoIn5,Zhou2013CeCoIn5, Neto2013, Rosenthal2014,Fan2015,Inoue2016}. Its ability to image electron scattering both in the occupied and unoccupied states with an energy resolution limited only by the temperature of the experiment (for normal metal tips) provides sufficient resolution to map out the structure of the superconducting order parameter\cite{hoffman_imaging_2002,wang_quasiparticle_2003}.
While the majority of works concentrated on determining the magnitude and $\bk$-space structure of the superconducting gap\cite{mcelroy_relating_2003, hanaguri_quasiparticle_2007, kohsaka_how_2008, Allan2012,Allan2013CeCoIn5,Zhou2013CeCoIn5}, QPI imaging carried out with and without magnetic field has been shown to provide phase sensitive information about the superconducting order parameter\cite{Hanaguri2009Dwave,Hanaguri2010,Fan2015}. The magnetic field produces vortices that act as additional scatterers contributing to the QPI, which increase the signal of certain scattering wave vectors. For the cuprate superconductors, relating these changes in the scattering intensity to the symmetry of the superconducting order parameter has worked quite well\cite{Hanaguri2009Dwave}, however for the iron chalcogenides, the interpretation of field-dependent QPI experiments has been disputed\cite{Hanaguri2010,Hanaguri2010comment,Hanaguri2010commentReply}. More recently, it has been argued that the interpretation of magnetic field dependent QPI data is not  straightforward\cite{HAEM2015}, in particular in cases where the vortex cores are more spatially extended objects, as happens in lower temperature superconductors. 
With a question mark on the general applicability of field-dependent QPI to study the symmetry of the superconducting condensate, new methods to determine the symmetry of the order parameter based on characterizing the scattering phase at individual defects have been proposed\cite{HAEM2015}.

While QPI has typically been analyzed in a way that discards the phase, the scattering phase \textit{does} encode important information about both the scatterer and the superconducting condensate. For a superconductor with normal $s_{++}$ symmetry, the phase shift between the scattering pattern in the occupied and unoccupied states can be used to extract information about the scattering strength of a magnetic defect\cite{menard_coherent_2015}. However, for a superconductor with a sign-changing order parameter, there can be an additional contribution to the phase if a quasiparticle scatters between states with a different sign.  
Here we present a detailed account of defect-bound state QPI (DBS-QPI), a new probe for the symmetry of the superconducting order parameter. DBS-QPI uses spatially extended impurity bound states which reside inside the superconducting gap. We demonstrate its application to LiFeAs \cite{ChiPRL2017}, and discuss here in detail its robustness for different models and its relation to other methods based on using differential conductance maps to determine the phase of the order parameter. 
To this end, we will use simulated scattering patterns for magnetic and non-magnetic impurities in $s_{++}$ and $s_\pm$ superconductors, comparing between different models and data. 

We will first introduce the theoretical framework to study the impact of defect scattering on the density of states, as well as the physical observables we will use to study the scattering phase (section II). This includes a discussion of the phase-referenced QPI which we have introduced as a means to analyze the DBS-QPI. We will then use these to simulate the QPI patterns from defect bound states, in particular considering the sign (phase) of the QPI amplitude, and demonstrate that consistent results are obtained for different models for the band structure (section III and Appendix A). In section IV, we compare the ratio map QPI, phase referenced DBS-QPI, and the HAEM method to analyze scattering patterns near defect bound states, discussing the strengths and weaknesses of each of these methods.

\section{\label{sec:theory} Scattering Theory and Phase in QPI}
\subsection{\label{subsec:theoryTmatrix} $T$-matrix approximation}

To study scattering by a defect, we use the following Hamiltonian: 
 \begin{equation}
 H = H_{0} + H_{\mrm{imp}},
 \label{eq:Hk}
 \end{equation}
where $H_0$ is the bare Hamiltonian including the superconducting BCS term, 
\begin{align}
H_{0} = 
\begin{bmatrix}
\epsilon(\bk) \hfill & \Delta(\bk) \\
\Delta^{\dagger}(\bk) & -\epsilon^{\mathrm{T}}(-\bk) \hfill 
\end{bmatrix},
\label{eq:H0}
\end{align}
with $\epsilon(\bk)$ representing the normal state band structure, and $\Delta(\bk)$ the superconducting order parameter. The impurity Hamiltonian for point-like defects, $H_{\mrm{imp}}$, is given by 
\begin{equation}
H_{\mrm{imp}} = V_{0}\sum_{\mu\sigma}c^{\dagger}_{0\mu\sigma}c_{0\mu\sigma}.
\label{eq:Hk_imp}
\end{equation}
It describes the scattering for orbital $\mu$ and spin $\sigma$ at site $\br$ = (0, 0) with potential $V_{0}$. Only intra-orbital scattering is considered in the calculation because it leads to the dominant components in the scattering matrix~\cite{GastiasoroPRB2013}. An equal strength of the scattering potential is assumed for all orbitals. In momentum space the potential is constant, given by 
\begin{align}
V(\bk,\bk^{\prime}) \equiv V = \frac{1}{N}V_{0}
\begin{bmatrix}
I \hfill & 0 \\
0 & \mp I \hfill 
\end{bmatrix},
\label{eq:Vkk} 
\end{align}
where $N$ is the system size, and the choice of the sign, - or +, is for nonmagnetic and magnetic defects, respectively.

Using the $T$-matrix approximation~\cite{Balatsky2006review}, the Green's function in the presence of a defect is given by
\begin{equation}
 G(\bk,\bk', \omega) = G_{0}(\bk, \omega)\delta_{\bk,\bk'} +  G_{0}(\bk, \omega)T_{\bk,\bk^{\prime}}(\omega) G_{0}(\bk', \omega),
\label{eq:Gk}
\end{equation}
where $G_{0} = [(\omega  + \mathrm{i}\eta)I - H_{0}]^{-1}$ and the $T$-matrix is given by 
\begin{equation}
 T_{\bk,\bk^{\prime}}(\omega) \equiv T(\omega) = [I - Vg(\omega)]^{-1}V,
\label{eq:Tw}
\end{equation}
with $g(\omega) = \sum_{\bk}G_{0}(\bk, \omega)$. The total density of states (DOS) $\tilde\rho(\bk,\omega)$ is obtained from the imaginary part of the Green's function
\begin{eqnarray}
\tilde\rho(\bk,\omega) &=& -\frac{\mrm{1}}{\pi} \mrm{Im} G(\bk,\bk, \omega)\nonumber \\ &=& -\frac{\mrm{1}}{\pi}\mrm{Im}G_{0}(\bk, \omega) -\frac{\mrm{1}}{\pi} \mrm{Im} G_{0}(\bk, \omega)T(\omega) G_{0}(\bk, \omega) \nonumber \\
&\equiv& \tilde\rho_0(\bk,\omega) + \delta\tilde\rho(\bk,\omega),
\label{eq:rhok}
\end{eqnarray}
where $\tilde\rho_0(\bk,\omega)$ is the bare DOS resulting from $H_0$, and $\delta\tilde\rho(\bk,\omega)$ is the perturbation of the DOS due to the presence of a defect.
  
QPI is calculated from the Fourier transform of the local density of states (LDOS) in real space. Given a defect at site $\br = (0,0)$， the Green's function in real space can be obtained from the $T$-matrix
\begin{equation}
 G(\br,\br', \omega) = G_{0}(\br,\br', \omega) +  NG_{0}(\br,{\bf 0}, \omega)T(\omega) G_{0}({\bf 0},\br', \omega),
\label{eq:Grr}
\end{equation}
so that the LDOS is given by
\begin{eqnarray}
 \rho(\br, \omega) &=& -\frac{\mrm{1}}{\pi} \mrm{Im} G(\br,\br, \omega) \nonumber \\
 &=& -\frac{\mrm{1}}{\pi}\mrm{Im}G_{0}(\br,\br, \omega)  -\frac{\mrm{1}}{\pi}\mrm{Im}NG_{0}(\br,{\bf 0}, \omega)T(\omega) G_{0}({\bf 0},\br, \omega) \nonumber \\
 &=& \rho_{0}(\br,\omega) + \delta \rho(\br,\omega).
\label{eq:rhorr}
\end{eqnarray}
Finally, the QPI map is given by 
\begin{equation}
\delta \tilde\rho(\bq, \omega) = \frac{1}{N}\sum_{\br}e^{-i\bq\cdot\br}\delta \rho(\br,\omega).
\label{eq:rhoq}
\end{equation}
In experiments, one measures the tunneling conductance map, $g(\br, \omega)$, which, if matrix element effects can be neglected, is proportional to $ \rho(\br, \omega)$.~\cite{STMTheory_Tersoff}

\subsection{\label{subsec:theoryReQPI} The phase in Fourier transformed QPI}
The Fourier transform of $g(\br, \omega)$  gives a map of complex values, $\tilde g(\bq, \omega)$, with real and imaginary parts. Typically, only the modulus of $\tilde g(\bq, \omega)$ is analyzed and the phase is ignored because the main interest is in identifying dominant scattering vectors $\bq$, i.e. where $|\tilde g(\bq,\omega)|$ has maxima. However, for the present purpose, the phase contains important information to discriminate the in-phase and antiphase signals between two energies. For a particular scattering vector $\bq$ at two energies $\omega_1$ and $\omega_2$, in-phase refers to $\tilde g(\bq, \omega_1) \propto \tilde g(\bq, \omega_2)e^{2n\mrm{\pi i}} = \tilde g(\bq, \omega_2)$, and anti-phase refers to $\tilde g(\bq, \omega_1) \propto \tilde g(\bq, \omega_2)e^{(2n+1)\mrm{\pi i}} = -\tilde g(\bq, \omega_2)$. In this study, the phase information $\tilde g(\bq, \omega)$ is preserved, allowing for the study of effects of different order parameters and defects of different nature. 

Defect apparent shape and size can affect the phase of the Fourier transform. First, the symmetry determines the signal weighting between the real and imagine parts. It is often assumed that defects have point symmetry, namely $g(\br,\omega)$ = $g(-\br,\omega)$ for a defect at the origin. With point symmetry of the scattering pattern, it can be shown that $\tilde g(\bf q,\omega)$ is real, because for the complex conjugate $\tilde g^\dagger(\bf q,\omega)$, we have
\begin{eqnarray}
\tilde g^{\dagger}(\bq, \omega) &=& \frac{1}{N}\sum_{\br}e^{i\bq\cdot\br} g(\br,\omega) \nonumber\\
					&=&\frac{1}{N}\sum_{\br}e^{-i\bq\cdot\br} g(-\br,\omega) \nonumber \\
                    &=&\frac{1}{N}\sum_{\br}e^{-i\bq\cdot\br} g(\br,\omega) \nonumber \\
                    &=& \tilde g(\bq, \omega).
\end{eqnarray}
Therefore, the QPI signals are all in the real part of $\tilde g(\bq, \omega)$. For defects without point symmetry, QPI signals are shared between the real part Re$[\tilde g(\bq, \omega)]$ and the imaginary part Imag$[\tilde g(\bq, \omega)]$. Second, the spatial shape of a defect can also affect the phase. Real space LDOS oscillations are shifted out from the defect center by the size of the defect, which contributes an additional phase shift in the Fourier transform. 

In theoretical calculations, the $\delta$ scattering potential is typically assumed to have point symmetry and hence the simulated QPI signals only exist in Re$[\tilde g(\bq, \omega)]$. It is also an ideal potential with zero size in $\br$-space. In experimental measurements, real defects have finite sizes and various shapes, which complicates the direct comparison between theory and experiments. Below, we discuss a new approach that can be taken to deal with phase in the Fourier transform of the tunneling conductance.

\subsection{\label{subsec:theoryPhaseref} Phase-referenced QPI}
In addition to complications in the phase arising from the symmetry, shape and size of a defect, the Fourier transform also includes an overall phase factor related to the defect positions in an image. Unless one is dealing with an ideal point scatterer situated at the centre of an image, all of these factors come into play in the complex Fourier transform of the tunneling conductance. Phase-referenced QPI has been introduced as a way to deal with this by zeroing the phase at positive energies and then applying this adjustment at negative energies.~\cite{ChiPRL2017}. Since the phase contributions coming from the nature and placement of the defects will generally contribute in the same way at positive and negative energies, this referencing of the phase leaves the contrast associated with a sign-changing order parameter in the negative energy images. This then makes it much more straightforward to compare the relative phase between QPI at positive and negative energies $\pm E$.
The Fourier transform of the tunneling conductance can be written as $|\tilde g(\mathrm{\textbf{q}}, E)|\times e^{i\theta_{\bq, E}}$, where $|\tilde g(\mathrm{\textbf{q}}, E)|$ is the intensity and $\theta_{\bq, E}$ is the phase at wave vector $\bq$ and energy $E$. A phase-referenced Fourier transform (PRFT) is obtained by taking the Fourier transform of $g(\mathrm{\textbf{r}}, E)$ at positive energy $E$, obtaining the phase factor $e^{i\theta_{\bq, E}}$, and then using that as a reference for the Fourier transform at negative energy $-E$. The PRFT of the tunneling conductance $\tilde g_c(\mathrm{\textbf{q}}, \pm E)$ for $E>0$ is given by 
\begin{eqnarray}
\tilde g_c(\mathrm{\textbf{q}}, E) &=& \mathrm{Re}(|\tilde g(\mathrm{\textbf{q}}, E)|\times e^{i\theta_{\bq, E}})\times e^{-i\theta_{\bq, E}} \nonumber \\
&=& |\tilde g(\mathrm{\textbf{q}}, E)| \\
\tilde g_c(\mathrm{\textbf{q}}, -E) &=& \mathrm{Re}\ (|\tilde g(\mathrm{\textbf{q}}, -E)|\times e^{i\theta_{\bq, -E}})\times e^{-i\theta_{\bq, E}} \nonumber \\
&=& |\tilde g(\mathrm{\textbf{q}}, -E)| \times \mathrm{Re}(e^{i(\theta_{\bq, -E}-\theta_{\bq, E})})  .
\label{g_c}
\end{eqnarray}
The phase factor $\mathrm{Re}(e^{i(\theta_{\bq, -E}-\theta_{\bq, E})})$ of the PRFT is $+1$ for in-phase oscillations, and $-1$ for out-of-phase oscillations. The PRFT is at the basis of DBS-QPI, enabling meaningful phase information to be extracted.

\section{Simulations of DBS-QPI} 
In this section, we present and discuss simulated DBS-QPI maps to demonstrate how information about the scatterer and the superconducting order parameter can be extracted.
We use the five-orbital tight-binding model for LiFeAs from Ref.~\onlinecite{GastiasoroPRB2013} for simulating QPI. The corresponding Fermi surface is shown in Fig.~1b. For the superconducting gap amplitude we use $\Delta(\bk) = \Delta_0 \cos(k_x)\cos(k_y)$ for the $s_{\pm}$ order parameter, and $|\Delta(\bk)| = | \Delta_0 \cos(k_x)\cos(k_y)|$ for the $s_{++}$ order parameter, respectively. Here we set $\Delta_0 = 0.016$~eV which is consistent with both the gap amplitude obtained for spin-fluctuation mediated pairing~\cite{GastiasoroPRB2013}, and with the experimental gap amplitudes using a band renormalization factor of 2-4~\cite{ChiPRB2016, ARPES_Umezawa,Allan2012, Hanaguri2012}. Both nonmagnetic and magnetic defects can generate in-gap bound states for an $s_{\pm}$ order parameter. However, only a magnetic defect can give rise to in-gap bound states in a superconductor with an $s_{++}$ order parameter. Here we have calculated DBS-QPI for all three scenarios that generate in-gap bound states.

\subsection{\label{subsec:QPIevctor} Identification of QPI vectors}

\begin{figure}[ht]
\centering
\includegraphics[width=0.79\linewidth]{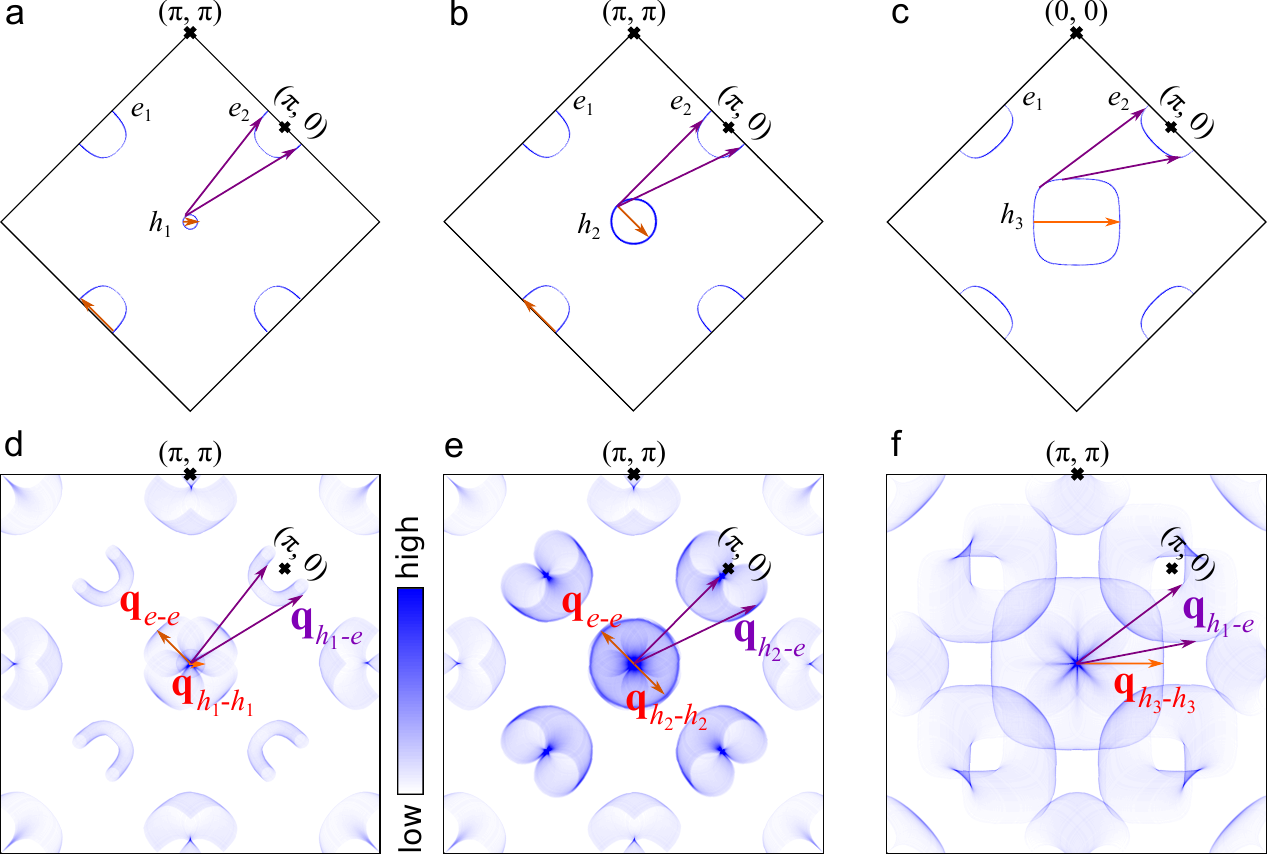}
\caption{\label{figS1_QPIVector} 
\textbf{Identification of QPI vectors in LiFeAs.} (\textbf{a})-(\textbf{c}) The isolated Fermi surfaces for [$h_1$, $e$], [$h_2$, $e$], and [$h_3$, $e$], respectively. In (\textbf{c}), the Brillouin zone is centered at $\bk = (\pi,\pi)$. (\textbf{d})-(\textbf{e}) The autocorrelation is given for each isolated Fermi surface of the upper panel.  
}
\end{figure}

For the case of five bands crossing $E_{\mathrm F}$, considerable care is required to identify intra-band and inter-band QPI features among all the available states. In the five-orbital model, we have separated the bands at $E_{\mathrm F}$ into three sets, with one hole band per set, and plotted their autocorrelations in Figure~\ref{figS1_QPIVector}. In this way, the origin of the intra-band and inter-band QPI features can be identified with confidence. In particular, from Figure~\ref{figS1_QPIVector}d-f, we are able to unambiguously separate the QPI features for the inter-band scattering between $h_1 - e$, $h_2 - e$ and $h_3 - e$. Here $e_1$ and $e_2$ are identical except for being in orthogonal directions, hence the QPI features are identical apart from a rotation by $\pi/2$. Therefore, the labels for electron bands have been omitted.

\subsection{\label{subsec:CalQPI} Comparison between QPI with and without superconductivity at $|E| > |\Delta|$}

\begin{figure}[ht]
\centering
\includegraphics[width=0.49\linewidth]{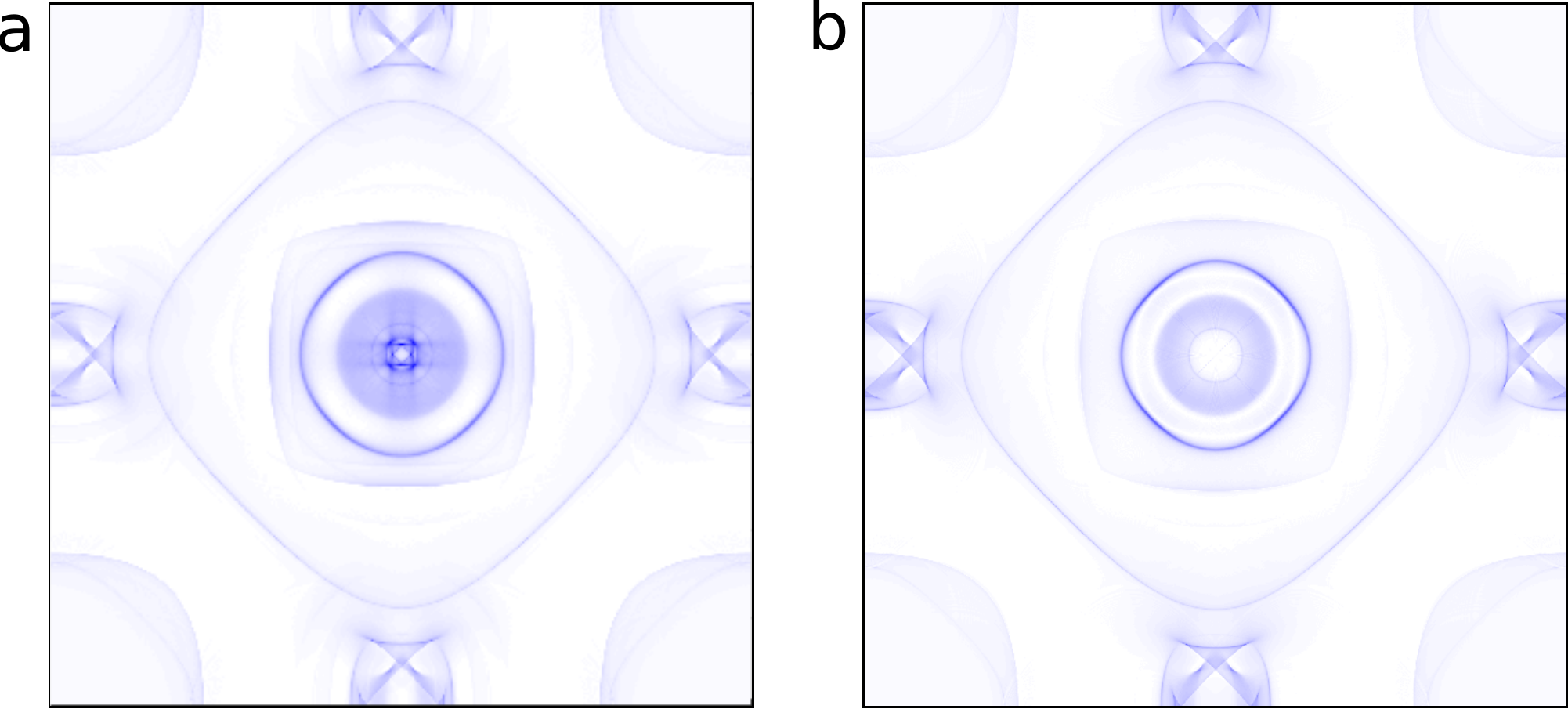}
\caption{\label{figS2_QPI_Normal_vs_SC} 
\textbf{QPI: superconducting state vs. normal state at $E = 1.3\Delta_{0}$.} (\textbf{a}) QPI in the superconducting state. (\textbf{b}) QPI in the normal state.   
}
\end{figure}

When $|E| > |\Delta|$, the superconducting coherence factors $u(\bk, E)$ and $v(\bk, E)$ approach their normal state values. Therefore, the QPI features in the superconducting state and the normal-state should be almost identical to each other. Figure~\ref{figS2_QPI_Normal_vs_SC} shows the QPI intensity maps at $E = 1.3\Delta_{0}$ in both the superconducting and the normal states. The two QPI maps agree very well on a quantitative level. Thus, QPI in the superconducting state at $|E| > |\Delta|$ is a good representation of QPI in the normal state.

\subsection{\label{subsec:Dispersion} DOS in momentum space}
\begin{figure}[h!]
\centering
\includegraphics[width=0.5\linewidth]{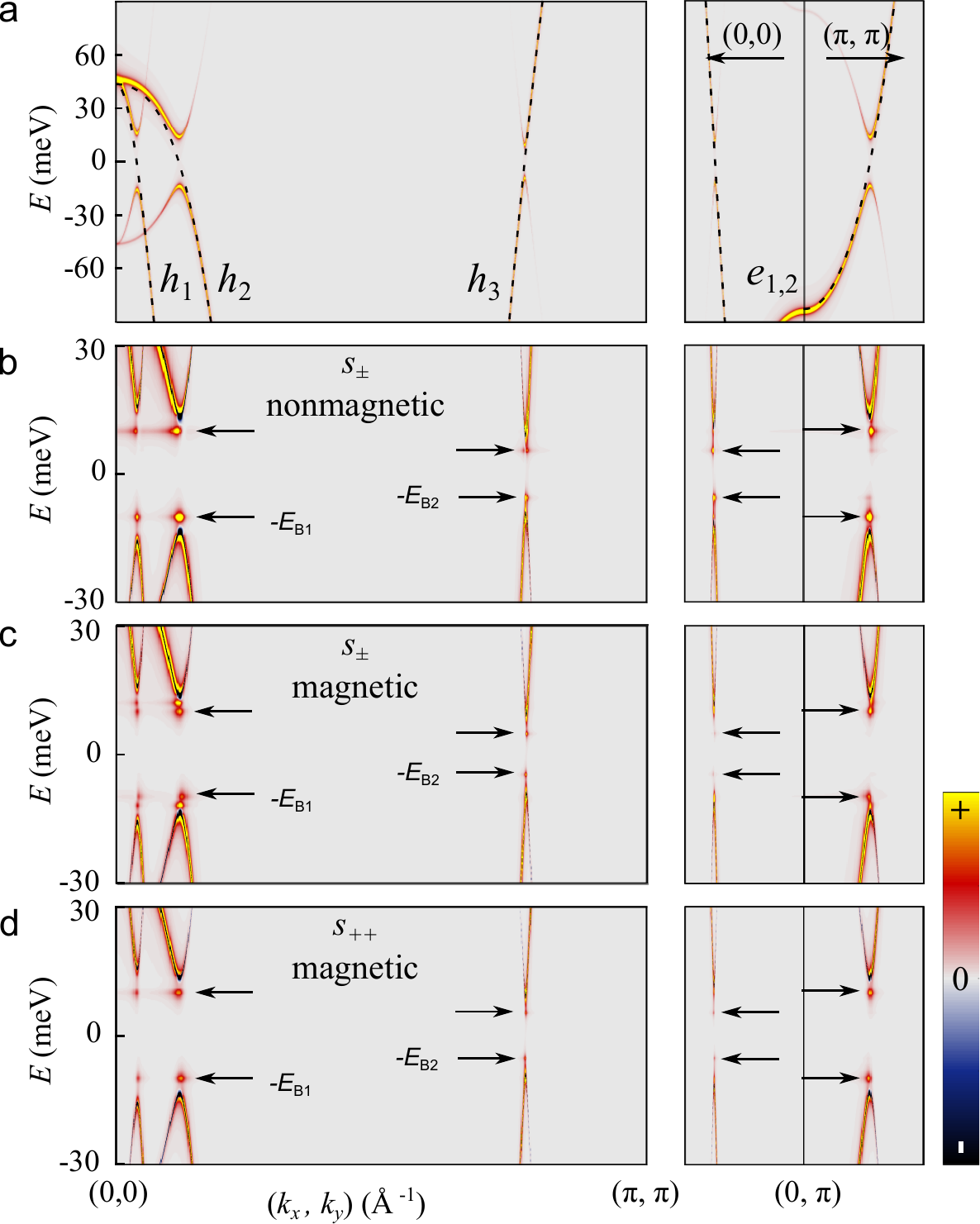}
\caption{\label{figS3_Dispersion} 
\textbf{DOS in $\bk$-space} (\textbf{a}) Superconducting bare DOS, $\rho_0(\bk,\omega)$, along with overlay of band dispersion (dashed line). (\textbf{b})-(\textbf{d}) $\delta\rho(\bk,\omega)$ for $s_{\pm}$ with nonmagnetic and magnetic defects and $s_{++}$  with a magnetic defect, respectively.   
}
\end{figure}

From Equation~\ref{eq:rhok}, we can calculate the bare DOS $\rho_0(\bk,\omega)$ and the change in DOS $\delta\rho(\bk,\omega)$ due to defect scattering. As shown in Figure~\ref{figS3_Dispersion}(a), the superconducting gaps open at $E_\mrm{F}$, with the large gaps in the bands of $h_{1}, h_{2}$, and $e$ along the $(0,\pi)$-$(\pi, \pi)$ direction and the small gaps in the bands of $h_{3}$ and $e$ along the $(0,\pi)$-$(0,0)$ direction,  consistent with experimental observations~\cite{ARPES_Umezawa}. 

Figure~\ref{figS3_Dispersion}(b)-(d) shows $\delta\rho(\bk,\omega)$ under three conditions that allow in-gap bound states. Scattering potentials were chosen to generate bound states with the bound state energy $E_{\mrm{B}1}$ close to the gap edge of the small gap, to imitate the measured bound state on an Fe-D$_2$ defect~\cite{ChiPRB2016}. The scattering potentials are $-1.3$~eV for $s_{\pm}$ with a nonmagnetic defect, $-0.6$~eV for $s_{\pm}$ with a magnetic defect, and $-0.35$~eV for $s_{++}$ with a nonmagnetic defect. The potential strength for $s_{\pm}$ with a nonmagnetic defect is consistent with the theoretically estimated value for native defects~\cite{KreiselPRB2016,ChiPRB2016}.  $\delta\rho(\bk,\omega)$ has both positive and negative values with $E > \Delta_i$, stemming from the slight spectral weight shift attributed to the defect potentials. However, $\delta\rho(\bk,\omega)$ is orders of magnitude smaller than $\rho_0(\bk,\omega)$ and hence the total DOS, $\rho(\bk,\omega)$, is positive for all states. For all three parameters in the calculation, bound states at large energy $E_{\mrm{B}1}$ are associated with the large gaps and bound states at small energy $E_{\mrm{B}2}$ are associated with the small gaps. In addition, all bound states follow the band dispersion and are confined in small momentum regions. These bound states consist of Bogoliubov quasiparticles, the excitations of Cooper pairs, and QPI can be measured from these Bogoliubov quasiparticles at the bound state energies.


\subsection{DBS-QPI without phase referencing}

\begin{figure}[ht!]
\centering
\includegraphics[width=0.77\linewidth]{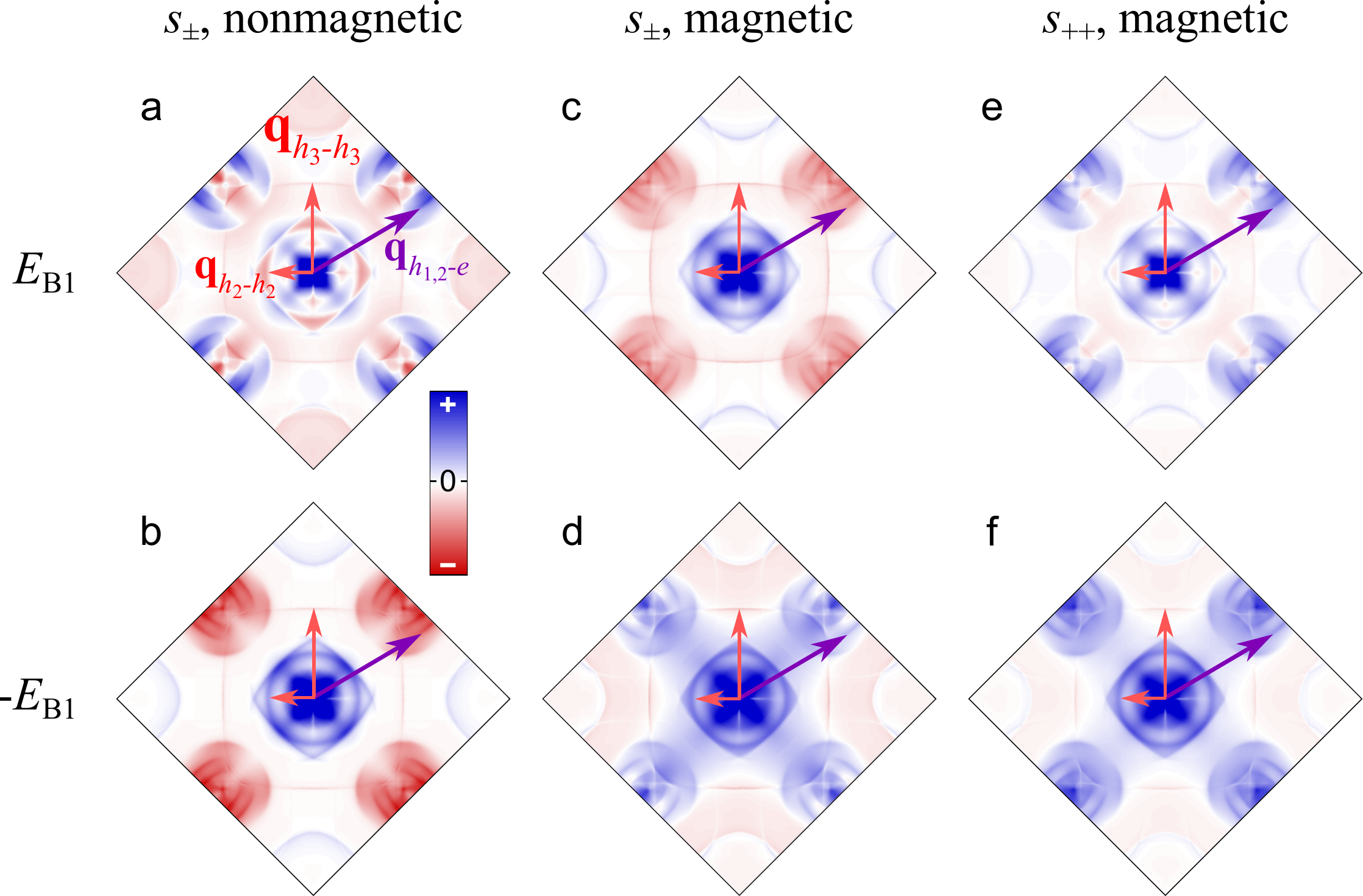}
\caption{\label{figS4_BoundStateEB1} 
\textbf{DBS-QPI at $\pm E_{\mrm{B1}}$.} Re[$\tilde g(\bq, \pm E_{\mrm{B}1})$] are shown for $s_{\pm}$ with a nonmagnetic potential and a magnetic potential, and for $s_{++}$ with a magnetic potential only. As opposed to $|\tilde g(\bq, \pm E_{\mrm{B}1})|$,  Re[$\tilde g(\bq, \pm E_{\mrm{B}1})$] reveals the phase of QPI features.    
}
\end{figure}

While the phase information in experimental data always suffers from the effects of global phase factors and details of the scatterers' size and symmetry, simulations using a point scatterer can yield meaningful phase information for DBS-QPI without any need for phase referencing. To highlight the effects that can appear in DBS-QPI due to a sign-changing superconducting order parameter, we discuss here the DBS-QPI without phase reference. Calculations employing the full phase-referenced DBS-QPI are shown and discussed in detail in Ref.~\onlinecite{ChiPRL2017} and are summarized below in section IV.

Figure~\ref{figS4_BoundStateEB1} shows the simulated DBS-QPI at $E_{\mrm{B}1}$ for all three combinations of order parameters and types of defects. $\tilde g(\bq, \pm E_{\mrm{B}1})$, which is proportional to $\tilde\delta \rho(\bq, \omega)$, was calculated according to Eq.~\ref{eq:rhoq}. Because an isotropic point-like potential at the origin is used in the calculation, all QPI signals are in the real part of $\tilde g(\bq, \omega)$ and the imaginary part is zero. Here, we show Re[$\tilde g(\bq, \pm E_{\mrm{B}1})$] instead of the absolute values $|\tilde g(\bq, \pm E_{\mrm{B}1})|$ in order to retain the phase information of the QPI oscillations after Fourier transformation. Comparing to Figure~\ref{figS1_QPIVector}, DBS-QPI at $E_{\mrm{B}1}$ mostly involves scattering within and between $h_{1,2}$ and $e$ bands as expected for the in-gap bound state of the large gaps. For all three cases, QPI signals are broadened at the bound states and the relative strength of the inter-band $\bq_{h_{1,2}-e}$ signal becomes enhanced (compare to Figure~\ref{figS2_QPI_Normal_vs_SC}). Thus, simply comparing the absolute value, $|\tilde g(\bq, \pm E_{\mrm{B}1})|$, is not adequate to distinguish between the order parameters $s_{\pm}$ and $s_{++}$. However, there are clear qualitative differences in the phase, i.e. in the regions of positive signal (blue) vs negative signal (red). For the phase-referenced DBS-QPI at $E_{\mrm{B}1}$, there is a sign inversion between $\pm E_{\mrm{B}1}$ for the majority of the $\bq_{h_{1,2}-e}$ signal in the cases of $s_{\pm}$ pairing with both nonmagnetic and magnetic scattering potentials. In contrast, the sign of $\bq_{h_{1,2}-e}$ scattering is mostly preserved between the positive and negative bound state energies. There are subtle quantitative distinctions between DBS-QPI of nonmagnetic and magnetic defects with the $s_{\pm}$ order parameter. However, the experimental data does not have adequate resolution to quantify these differences from the bound-state QPI measured at $E_{\mrm{B}1}$.


\begin{figure}[ht!]
\centering
\includegraphics[width=0.77\linewidth]{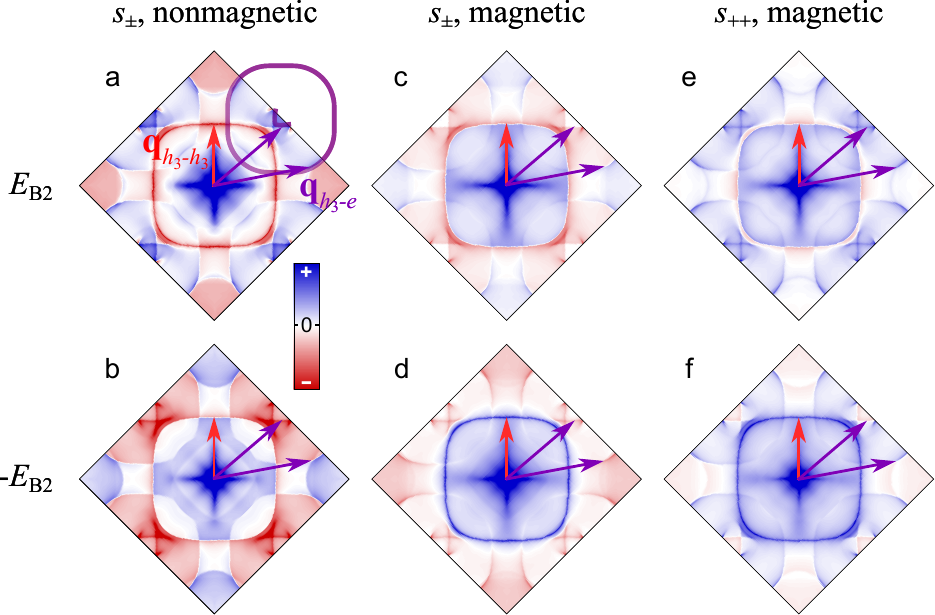}
\caption{\label{figS6_BoundStateEB2} 
\textbf{DBS-QPI at $\pm E_{\mrm{B2}}$.} DBS-QPI maps Re[$\tilde g(\bq, \pm E_{\mrm{B}2})$] are shown for $s_{\pm}$ with nonmagnetic and magnetic defects, and $s_{++}$ with a magnetic defect. The small $\bq_{h_3-e}$ arc right near ($\pi, 0$) is inward because the calculation is performed in the 1st Brillouin zone. It appears outward in the autocorrelation (see Figure~\ref{figS1_QPIVector}f) and experimental data (compare Figs.~4b and 4f of Ref.~\onlinecite{ChiPRL2017} and  Figure~\ref{fig5}a).    
}
\end{figure}

DBS-QPI at $E_{\mrm{B}2}$ are shown in  Figure~\ref{figS6_BoundStateEB2}. Only QPI features within and between $h_3$ and $e$ bands appear for this set of bound states, consistent with the bound states being attributed to the bands with small gaps. In the Re[$\tilde g(\bq, \pm E_{\mrm{B}2})$] images, the relative signal for inter-band scattering $\bq_{h_3-e}$ is strong for $s_{\pm}$ with a nonmagnetic defect, while the relative signal for intra-band scattering at $\bq_{h_3-h_3}$ is strong for $s_{\pm}$ with a magnetic defect, and $s_{++}$ with a magnetic defect. Moreover, at the wave vector $\bq_{h_3-e}$, the signal changes sign between $+E_{\mrm{B}2}$ and $-E_{\mrm{B}2}$ in the case of $s_{\pm}$ with a nonmagnetic defect. However, it only partially changes sign and mostly keeps the same phase between $\pm E_{\mrm{B}2}$ in the other two cases. This trend is opposite at the vector $\bq_{h_3-h_3}$. By looking at the ring corresponding to $\bq_{h_3-h_3}$ and its extension outward, it has negative signal (red) for both polarities at $E_{\mrm{B}2}$ for $s_{\pm}$ with a nonmagnetic defect, while it has the opposite sign between $\pm E_{\mrm{B}2}$ for the other two cases. Thus, by combining both sets of bound states, one is able to distinguish not only the order parameter ($s_{\pm}$) but also the nature of the defect (non-magnetic). These effects in DBS-QPI have also been tested using a two orbital model (Appendix A), which gives qualitatively consistent results, so the phase variation appears to be independent of the details of the band-structure model.


\section{\label{sec:expboundS} Three approaches to phase information in QPI data}
We apply three different approaches to extracting phase information from real QPI taken on LiFeAs (details of the experiment given in Appendix B). The data is the same as that used in Ref.~\onlinecite{ChiPRL2017}, where the phase-referenced approach was shown in detail. Here we contrast that approach with two other techniques: the ratio-map QPI introduced by Hanaguri et al.~\cite{hanaguri_quasiparticle_2007}, and the method developed by Hirschfeld, Altenfeld, Eremin and Mazin~\cite{HAEM2015}.

\subsection{Ratio-map DBS-QPI: Comparison between experimental data and calculations}

One method to analyze QPI data (and to some degree obtain information about the phase of QPI) is to use ratio-map QPI. This was initially introduced to reduce the in-phase systematic errors in  conductance maps due to the setpoint effect.\cite{hanaguri_quasiparticle_2007} The ratio map in real space is defined as:
\begin{equation}
Z(\mathrm{\textbf{r}}, \omega) \equiv \frac{g(\br, \omega)}{g(\br, -\omega)} = \frac{\rho(\br, \omega)}{\rho(\br, -\omega)}.
\label{eq:ratiomap}
\end{equation}
The ratio-map QPI, $|\tilde Z(\mathrm{\textbf{q}}, \omega)|$, is obtained by taking the absolute value of Fourier transformation of $Z(\mathrm{\textbf{r}}, \omega)$. In the ratio-map QPI, the in-phase scattering components at $\bq$ occuring at $+\omega$ and $-\omega$ are suppressed while antiphase components are enhanced. Therefore, the ratio-map QPI is a good way to examine the in-phase and anti-phase signals between two energies. Care is needed in interpreting it, since the ratio map can also provide misleading information about the phase change of the QPI signal. For example, if a modulation in the local density of states is in-phase but significantly stronger at $\omega$ than at $-\omega$, the ratio map barely cancels the in-phase part and enhances the anti-phase part. However, with this caveat in mind, ratio-map QPI is a complementary technique that can be used to enhance scattering signals which are out-of-phase as identified by phase-referenced QPI. Similar to the phase-referenced QPI method, it also is only applicable if the change of scattering vectors due to the normal state band dispersion is negligible, which is indeed the case for the energy range of interest in LiFeAs.

\begin{figure}[ht]
\centering
\includegraphics[width=0.77\linewidth]{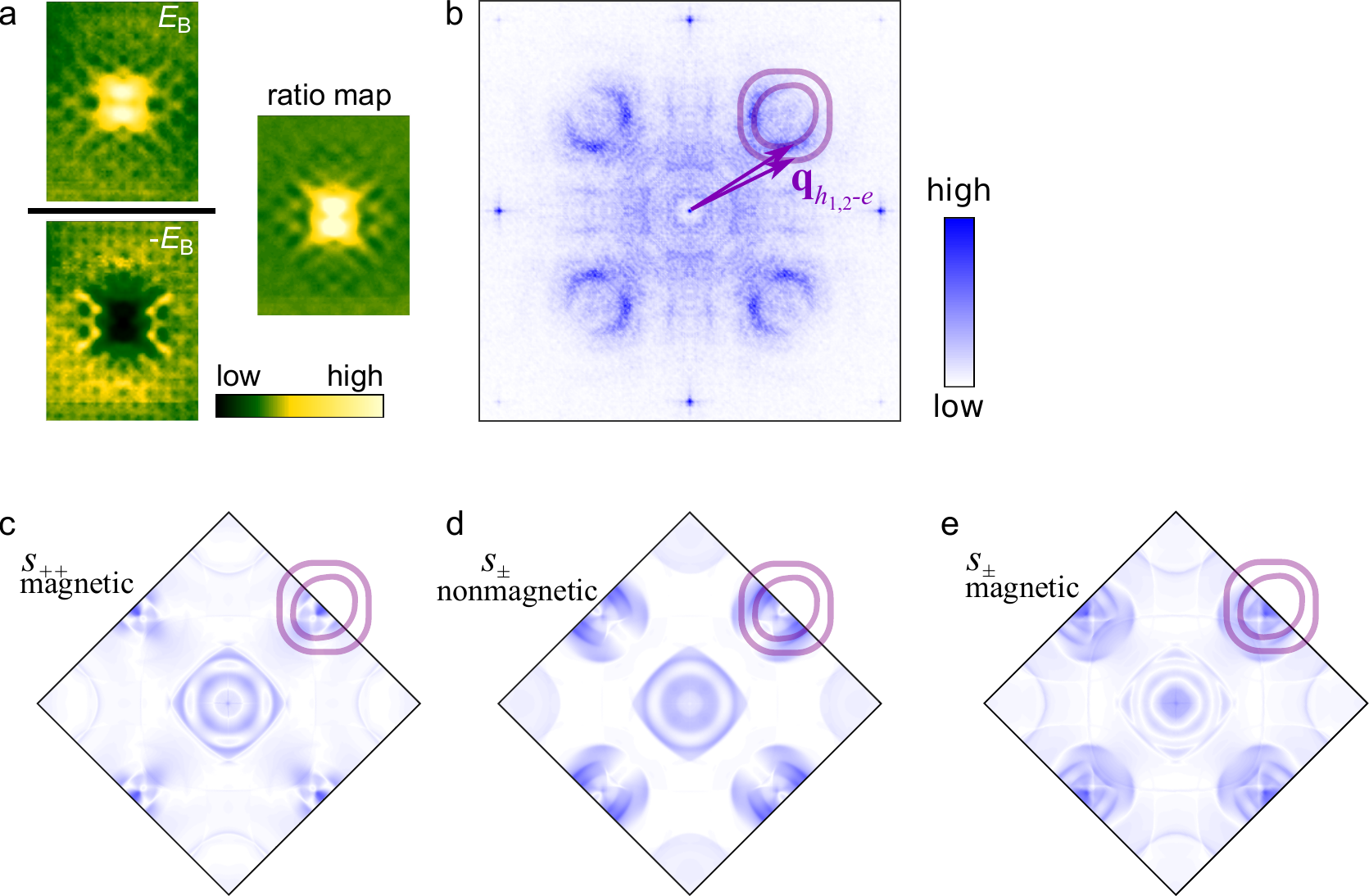}
\caption{\label{fig4} 
\textbf{Ratio-map DBS-QPI associated with the large gaps.} (\textbf{a}) The QPI oscillations near an Fe-D$_2$ defect at the bound state energies and the ratio map $Z(\mathrm{\textbf{r}}, E_{\mrm{B1}}) = g(\mathrm{\textbf{r}}, +E_{\mrm{B1}})/g(\mathrm{\textbf{r}}, -E_{\mrm{B1}})$. (\textbf{b}) The ratio-map DBS-QPI $|\tilde Z(\mathrm{\textbf{q}}, E_{\mrm{B1}})|$ obtained by Fourier transforming the ratio map $Z(\mathrm{\textbf{r}}, E_{\mrm{B1}})$. The $\bq_{h_{1,2}-e}$ QPI features (the ovals centered at $(0, \pi)$) is enhanced and the rest of QPI features are suppressed. (\textbf{c})-(\textbf{e}) The simulated ratio DBS-QPI with the settings of $s_{++}$ and a magnetic potential, $s_{\pm}$ and a nonmagnetic potential, and $s_{\pm}$ and a magnetic potential. 
}
\end{figure}

Figure~\ref{fig4}a shows the $\br$-space LDOS oscillations near an Fe-D$_2$ defect at $\pm E_{\mrm{B1}}$ and their ratio map, $Z(\mathrm{\textbf{r}}, E_{\mrm{B1}})$. Oscillations can be clearly seen near the defect in the ratio map, confirming the existence of strong anti-phase signals between $\pm E_{B1}$. By Fourier transforming $Z(\mathrm{\textbf{r}} , E_{\mrm{B1}})$ for the large scale image shown in Appendix B, $|\tilde Z(\mathrm{\textbf{q}}, E_{\mrm{B1}})|$ is obtained and is shown in  Figure~\ref{fig4}b. Comparing to QPI at the bound state energies (Fig.~3c and 3d in Ref.~\onlinecite{ChiPRL2017}), the $\bq_{h_{1,2}-e}$ features (highlighted by the purple ovals) become enhanced relative to the other QPI features, consistent with the sign change of the order parameter for inter-band scattering wavevectors $\bq_{h_{1,2}-e}$ in the phase-referenced DBS-QPI.  Figures~\ref{fig4}c-e show the simulations of ratio-map DBS-QPI in the three cases that produce in-gap bound states. The simulations based on $s_{\pm}$ agree best with experimental observations, consistent with the results from phase-referenced DBS-QPI.

\begin{figure}[ht!]
\centering
\includegraphics[width=0.99\linewidth]{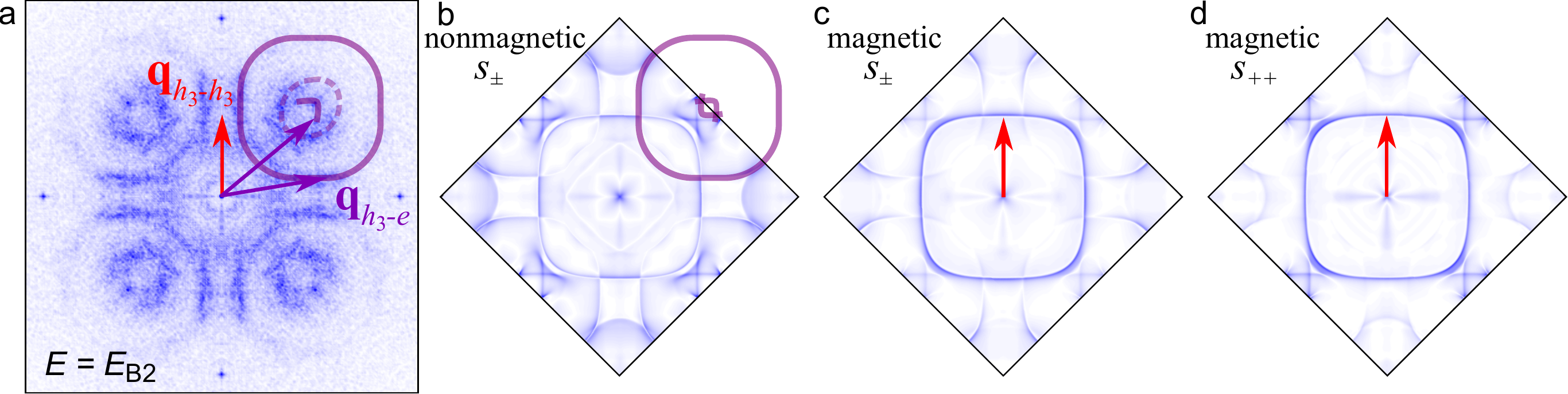}
\caption{\label{fig5} 
\textbf{Ratio-map DBS-QPI associated with the small gaps.} (\textbf{a}) Measured $|\tilde Z(\bq, E_{\mrm{B2}})|$ with $E_{\mrm{B2}} = 1.2~\mathrm{meV}$. Three QPI features centered at $(0, \pi)$ are highlighted by the arc and ovals. The position of $h_3$-$h_3$ scattering is indicated by the red arrow where signal is minimal. (\textbf{b})-(\textbf{d}) Simulated $|\tilde Z(\bq, E_{\mrm{B2}})|$. The $\bq_{h_3-e}$ and $\bq_{h_3-h_3}$ QPI features are indicated by the purple shapes and the red arrows, respectively. 
}
\end{figure}
 
The ratio-map method was applied to the bound states at $E_{\mathrm{B2}}$, as shown in  Figure~\ref{fig5}. In the experimental data, three inter-band QPI signals dominate, as indicated by the purple shapes. This agrees well with the sign change of $\bq_{h_{3}-e}$ observed in the phase-referenced DBS-QPI (see Fig.~4 in Ref.~\onlinecite{ChiPRL2017}). The middle oval (dashed shape) is the $\bq_{h_{1,2}-e}$ QPI features from $E_{\mrm{B}1}$ and present here because of thermal broadening effects. The other two shapes are $\bq_{h_{3}-e}$ QPI features from scattering between in-gap bound states for the small gaps in $h_3$ and $e$ bands. QPI features for intra-band $\bq_{h_3-h_3}$ were not well resolved within our measurement resolution (red arrow in  Figure~\ref{fig5}a) due to the in-phase cancellation effect. The simulated ratio-map bound-state QPI are shown in  Figure~\ref{fig5}b-d for the three possible scenarios that allow in-gap bound states. The $\bq_{h_3-e}$ signal is enhanced in the simulation using $s_{\pm}$ with a nonmagnetic defect. However, the other two cases have the $\bq_{h_3-h_3}$ signal enhanced as indicated by the red arrows. Therefore, only the simulation using $s_{\pm}$ with a nonmagnetic defect reproduces the dominant signatures seen in experimental data.

\subsection{\label{sec:HAEMCompare} Phase information in the HAEM method}  
Hirschfeld, Altenfeld, Eremin, and Mazin (HAEM) proposed a method using QPI to determine the order parameter of iron-based superconductors~\cite{HAEM2015}, and this has recently been applied to FeSe~\cite{sprau2016FeSe}.
In the HAEM method, the quantity considered is 
\begin{equation}
\tilde g^{\pm}(\bq,\omega) = \mathrm{Re}\ \left[ \tilde g(\bq,\omega) \pm \tilde g(\bq,-\omega) \right],
\label{eq:HAEMrhoq_pm}
\end{equation}
where $\tilde g(\bq,\omega)$ is the QPI signal amplitude at energy $\omega$ and scattering vector $\bq$~\cite{HAEM2015,HAEM2017}. In fact, $\tilde g^{+}(\bq,\omega)$ enhances in-phase QPI features and suppresses anti-phase QPI features, whereas $\tilde g^{-}(\bq,\omega)$ enhances anti-phase QPI features and suppresses in-phase QPI features. For example, if there are anti-phase QPI features, one has 
\begin{equation}
g(\br,\omega) \approx -g(\br,-\omega).
\label{eq:HAEMrhor}
\end{equation}
The Fourier transform gives
\begin{equation}
\tilde g(\bq,\omega) \approx -\tilde g(\bq,-\omega).
\label{eq:HAEMrhoq}
\end{equation}
Therefore, the HAEM signal is given by
\begin{eqnarray}
\tilde g^{+}(\bq,\omega) &\approx & 0 \\
\tilde g^{-}(\bq,\omega) &\approx & 2g(\bq,\omega). 
\label{eq:HAEMrhoq_pm_result}
\end{eqnarray}
As a result, $\tilde g^{-}$ doubles the signal with anti-phase oscillations and cancels the signal with in-phase oscillations. Essentially, similar to ratio-map QPI, the HAEM method sets apart the in-phase and anti-phase signals. It also suffers the same problem that ratio-map QPI has, as discussed in the section above: in-phase signals that are very different in amplitudes at positive and negative energies can confuse the interpretation. In principle, the HAEM method should produce results that are consistent with the ratio-map QPI. 

Experimentally, there are a few advantages of phase-referenced QPI and ratio-map QPI over the HAEM method. First, phase-referenced QPI and the ratio maps take account of all defects in the measured area, providing better signal-to-noise ratio. Second, signal integration in the HAEM method is effective in theory using an ideal $\delta$-potential, but in experiments, defects have spatial form factors which shift oscillations. This leads to extra complexity in the phase of QPI at different scattering vectors $\bq$ after Fourier transformation. Integration over an area with signals of varying complex phase further reduces the signal-to-noise ratio, rendering comparison with theory difficult. 

\begin{figure}[ht]
\centering
\includegraphics[width=0.49\linewidth]{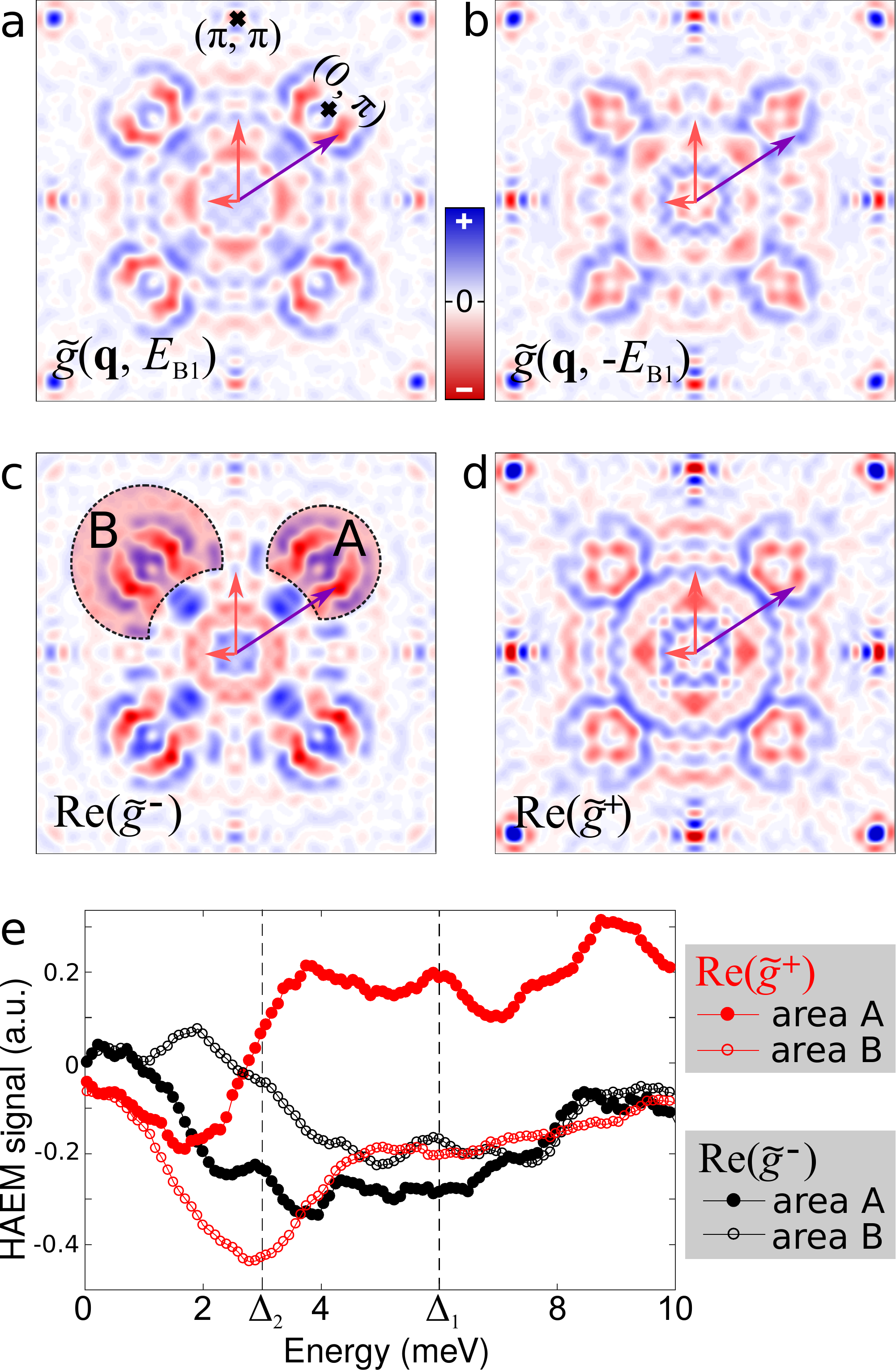}
\caption{\label{figS11_HAEM} 
\textbf{HAEM signal for a single Fe-D$_2$ defect.}  (\textbf{a}) $\tilde g(\bq, E_{\mrm{B1}})$ and (\textbf{b}) $\tilde g(\bq, -E_{\mrm{B1}})$ of the Fe-D$_2$ defect (after symmetrization and interpolation). (\textbf{c}) $\tilde g^{-}(\bq, E_{\mrm{B1}})$ and (\textbf{d}) $\tilde g^{+}(\bq, -E_{\mrm{B1}})$ of the Fe-D$_2$ defect. QPI maps in (\textbf{b})-(\textbf{e}) are in the same color scale. (\textbf{e}) The integrated inter-band $g^{-}$ and $g^{+}$ for the two areas indicated in (\textbf{d}). Here sample bias (mV) is converted to energy (meV).
}
\end{figure}

We applied the HAEM method to our experimental data for an Fe-D$_2$ defect. One relatively isolated defect was selected from the area in Figure~\ref{figS8_TopoSTS} with a size of $6.6\times 6.6$~nm$^2$ as indicated by the yellow square. The DBS-QPI of this defect is shown in Figure~\ref{figS11_HAEM}a and \ref{figS11_HAEM}b, and exhibits a complicated phase pattern after Fourier transformation. In the HAEM map, the anti-phase signal ($\tilde g^{-}$) primarily corresponds to $\bq_{h_{1,2}-e}$ (see  Figure~\ref{figS11_HAEM}c) and the in-phase signal does not exhibit scattering of significant strength (see  Figure~\ref{figS11_HAEM}d). $\tilde g^{-}(\omega)$ and $\tilde g^{+}(\omega)$ for $\bq_{h_{1,2}-e}$ was integrated over areas A and B, indicated in Figure~\ref{figS11_HAEM}c. For the two different integration areas, the shapes of $\tilde g^{-}(\omega)$ and $\tilde g^{+}(\omega)$ change on a scale comparable to the signal strengths themselves. This is because of the relatively complicated phase features for the $\bq_{h_{1,2}-e}$ QPI in LiFeAs, as well as the lower signal-to-noise ratio for a single defect. In contrast, the HAEM approach produced a much more robust result in work on FeSe~\cite{sprau2016FeSe} A likely source of the difference between the results on the two materials is that the apparent size of the defects in the topography of FeSe is smaller, making the influence of the defect form factor smaller in the QPI results on this material.

\subsection{\label{sec:Phasereferenced} Comparison to phase-referenced DBS-QPI} 

\begin{figure}[ht]
\centering
\includegraphics[width=0.49\linewidth]{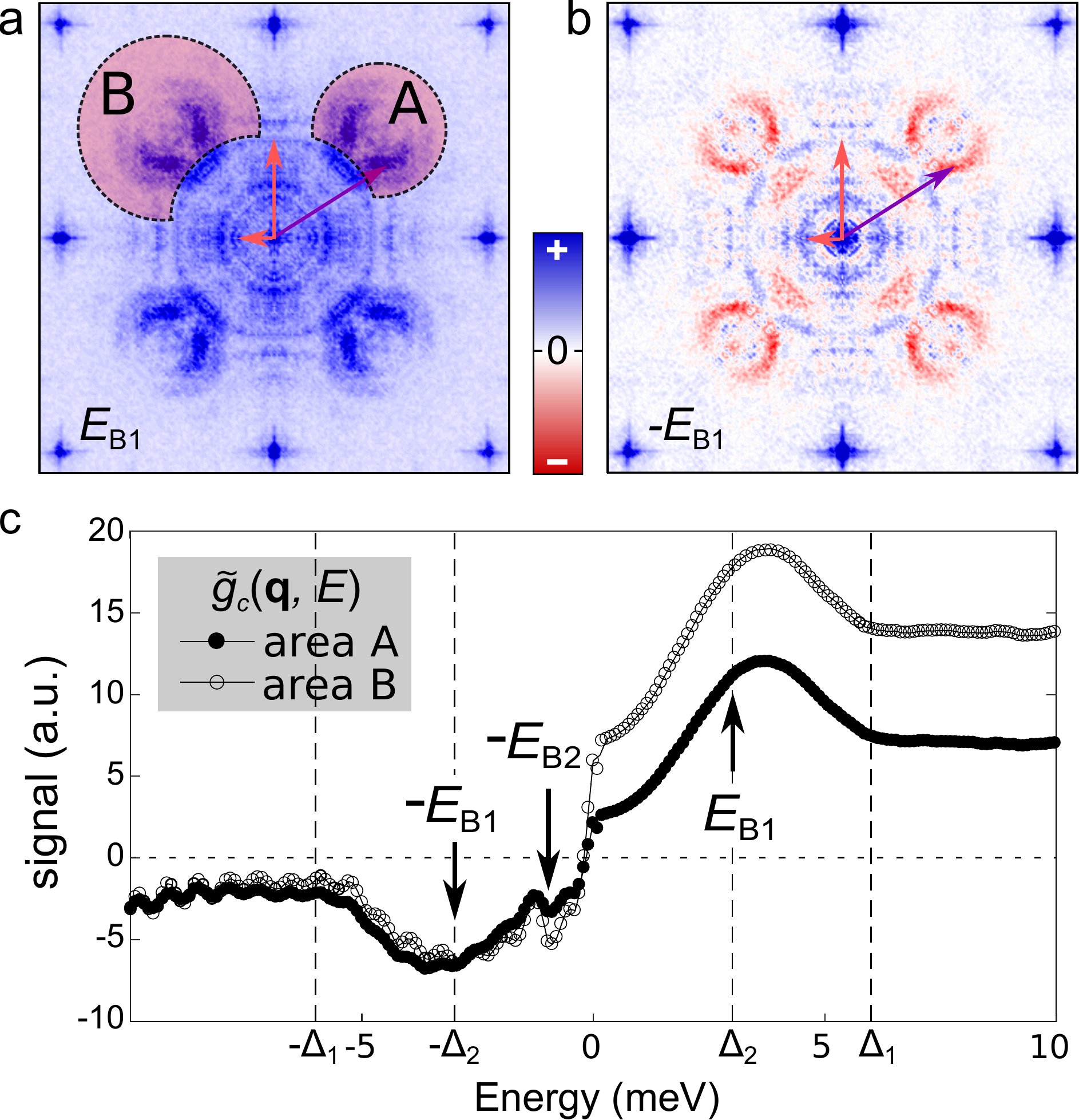}
\caption{\label{figS13_PRFT} 
\textbf{The integrated signal of Phase-referenced Fourier transform QPI.} (\textbf{a}) $\tilde g_c(\bq, E_{\mrm{B1}})$ and (\textbf{b}) $\tilde g_c(\bq, -E_{\mrm{B1}})$ including all defects. (\textbf{c}) The integrated inter-band $g_c(\bq, E)$ for the two areas indicated in (\textbf{a}).
}
\end{figure}

Figures~\ref{figS13_PRFT}a-b shows the DBS-QPI at $\pm E_{\mrm{B1}}$ after PRFT~\cite{ChiPRL2017}. The effect of the technique is apparent: the information in the form of sign contrast is moved to negative energies, and the sign difference between positive and negative energies for the interband scattering vectors is maximized. The inter-band QPI signal is integrated over the same two areas as those used above for the HAEM method. As shown in Figure~\ref{figS13_PRFT}c, the result is relatively independent of the integration windows, and the inter-band QPI signal is strongest near the bound state energies. The slight shift of the peak near $E_{\mrm{B1}}$ is due to thermal broadening at the $4.2$~K measurement temperature. With a larger integration window (area B) the overall amplitude is shifted upward because of more background noise being included, yet the overall shape remains the same. This demonstrates the robustness of PRFT with respect to the choice of the integration windows in the case of LiFeAs.

A small peak at -$E_{\mrm{B2}}$ for the larger integration window B can also be seen. This is because window B covers all the QPI signals of $\bq_{h_{1,2}-e}$ and $\bq_{h_3-e}$, while window A mainly covers $\bq_{h_{1,2}-e}$. It only shows up at negative energies for two reasons. First, $\bq_{h_3-e}$ is stronger at the negative energy (see Fig.4 in Ref.~\onlinecite{ChiPRL2017}). Second, the positive energy is zero-phased, so has contributions from the absolute values of both the QPI signal and the noise background. The weak signal at $E_{\mrm{B2}}$ is merged in the background noise integration. However, at negative energies, the background noise has random phase, and therefore the integration cancels the background noise significantly.

\section{Conclusion}

We have shown that by using the phase information contained in the spatial modulation of defect bound states, whose contribution to the QPI signal is usually ignored, information about the phase of the superconducting order parameter as well as the properties of the scatterer can be extracted. The robustness of the DBS-QPI method is demonstrated by comparison of a number of models against experimental data. To enable an experimentally robust analysis of the phase, we have introduced the phase-referenced DBS-QPI, which provides a systematic framework in which the phase of defect scattering can be interpreted. Our work provides a direct link between the theoretically predicted phase shifts for quasiparticle scattering at defects and experimental results, and provides strong evidence for an $s_{\pm}$ pairing state in LiFeAs.
\section{Acknowledgements}
The authors are grateful for helpful conversations with George Sawatzky, Mona Berciu, Peter Hirschfeld, Andreas Kreisel, and Steven Johnston. Research at UBC was supported by the Natural Sciences and Engineering Research Council, the Canadian Institute for Advanced Research, and the Canadian Foundation for Innovation. SAB was further supported by the Canada Research Chairs program.  PW acknowledges support from EPSRC grant no EP/I031014/1 and DFG SPP1458.

\appendix
\section{\label{sec:theorytwo} Simulations using a two orbital model}
\begin{figure}[h!]
\centering
\includegraphics[width=0.49\linewidth]{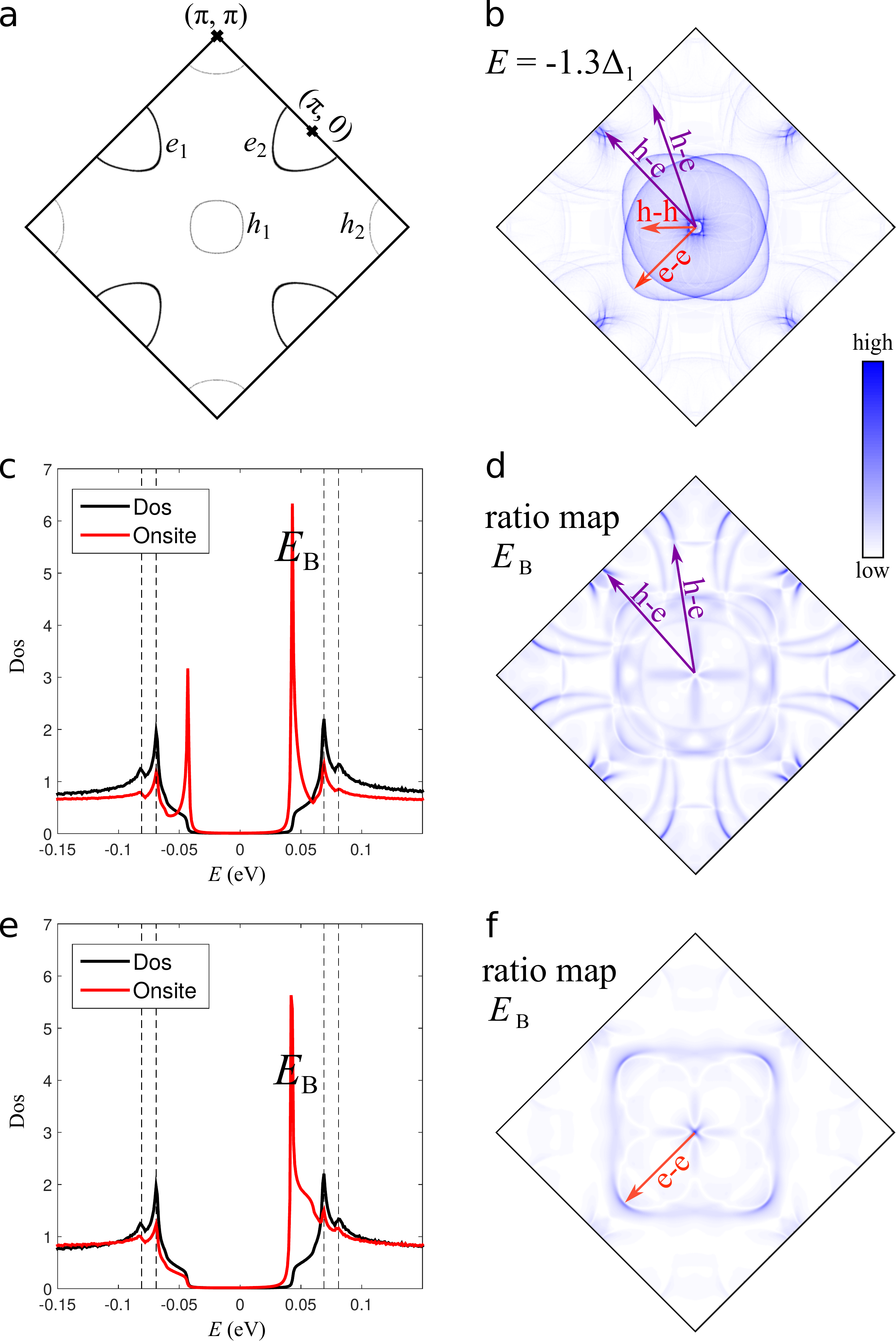}
\caption{\label{figS7_TwoOrbitals} 
\textbf{Bound state QPI from the two-orbital model.} (\textbf{a}) The Fermi surface of the two-orbital model. (\textbf{b}) $\tilde g(\bq, E = -1.3\Delta_0)$ QPI outside the superconducting gap. (\textbf{c}) LDOS without defect (black), and on the defect site (red), calculated with the input of $s_{\pm}$ and a nonmagnetic scattering potential. (\textbf{d}) The ratio-map QPI $\tilde Z(\bq, E_{\mrm{B}})$ of the bound state seen in (\textbf{c}). (\textbf{e}) LDOS without defect (black) and on the defect site (red) calculated with the input of $s_{++}$ and a magnetic scattering potential. (\textbf{f}) The ratio-map QPI $\tilde Z(\bq, E_{\mrm{B}2})$ of the bound state seen in (\textbf{e}). 
}
\end{figure}

To test the robustness of the conclusions reached from calculations within the five orbital model, we have tested the result using a minimal two-orbital model. We employ a model from Raghu \textit{et al.}~\cite{RaghuPRB2009_TwoOrb} to qualitatively compare bound state QPI between theory and experiment. Similar to the five-orbital model, the superconducting gap is $\Delta_\bk = \Delta_0 \cos(k_x)\cos(k_y)$ for $s_{\pm}$ order parameter and $|\Delta_{\bk}|$ for $s_{++}$ order parameter, respectively. $\Delta_0 = 0.09$~eV is used in the simulations; the precise value does not affect the qualitative result. In Figure~\ref{figS7_TwoOrbitals}(a) and (b), the Fermi surface of this model and its QPI for energies outside the superconducting gaps are shown. In this model, the DOS is dominated by the electron band, hence the prominent QPI intensities are intra-band $\bq_{e-e}$ and inter-band $\bq_{h-e}$ features. The intra-band $\bq_{h-h}$ QPI feature is very weak. In the superconducting state, two gaps open at $E_\mrm{F}$ as indicated by the dashed lines in Figure~\ref{figS7_TwoOrbitals}(c). Both a nonmagnetic defect ($V_0 = -1.8$~eV) with $s_{\pm}$ order parameter and a magnetic defect ($V_0 = -0.6$~eV) with $s_{++}$ order parameter produce strong in-gap bound states, as shown in  Figure~\ref{figS7_TwoOrbitals}(c) and (e). In their ratio-map QPI (see Figure~\ref{figS7_TwoOrbitals}(d) and (f)), the $\bq_{h-e}$ features are enhanced and $\bq_{e-e}$ features are suppressed in the simulation using the $s_{\pm}$  with a nonmagnetic defect. However, the opposite intensity changes occur in the simulation using $s_{++}$ with a magnetic defect. These results agree very well qualitatively with the results calculated from the five-orbital model.

\section{Supporting experimental data}
\begin{figure}[h!]
\centering
\includegraphics[width=0.37\linewidth]{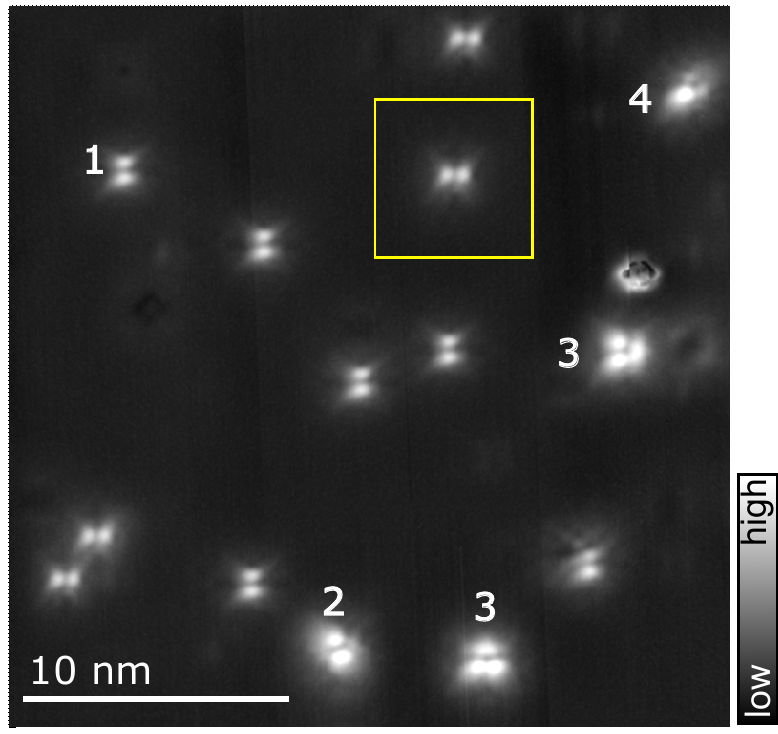}
\caption{\label{figS8_TopoSTS} 
\textbf{Topography of an area used for QPI measurements.} The types of native defects shown in this area are 1: Fe-D$_2$, 2: Fe-C$_2$, 3: Li-D$_1$, 4: As-D$_1$. The yellow square is a $6.6\times 6.6$~nm$^2$ area selected for the HAEM analysis.
}
\end{figure}
Differential conductance maps were acquired at a temperature of 4.2~K in a commercial Createc STM on single crystals of LiFeAs, cleaved at a temperature below 20~K\cite{ChiPRL2017}. Two conductance maps were measured for QPI. The results from the two datasets are consistent and the average of them is shown in this report.  Figure~\ref{figS8_TopoSTS} shows the topography of the area for one dataset. Taking one dataset as an example, there are fourteen native defects in the area of the sample: ten Fe-D$_2$ defects, one As-D$_1$ defect, two Li-D$_1$ defects, and one Fe-C$_2$ defect (Figure~\ref{figS8_TopoSTS}).

\begin{figure}[ht!]
\centering
\includegraphics[width=0.49\linewidth]{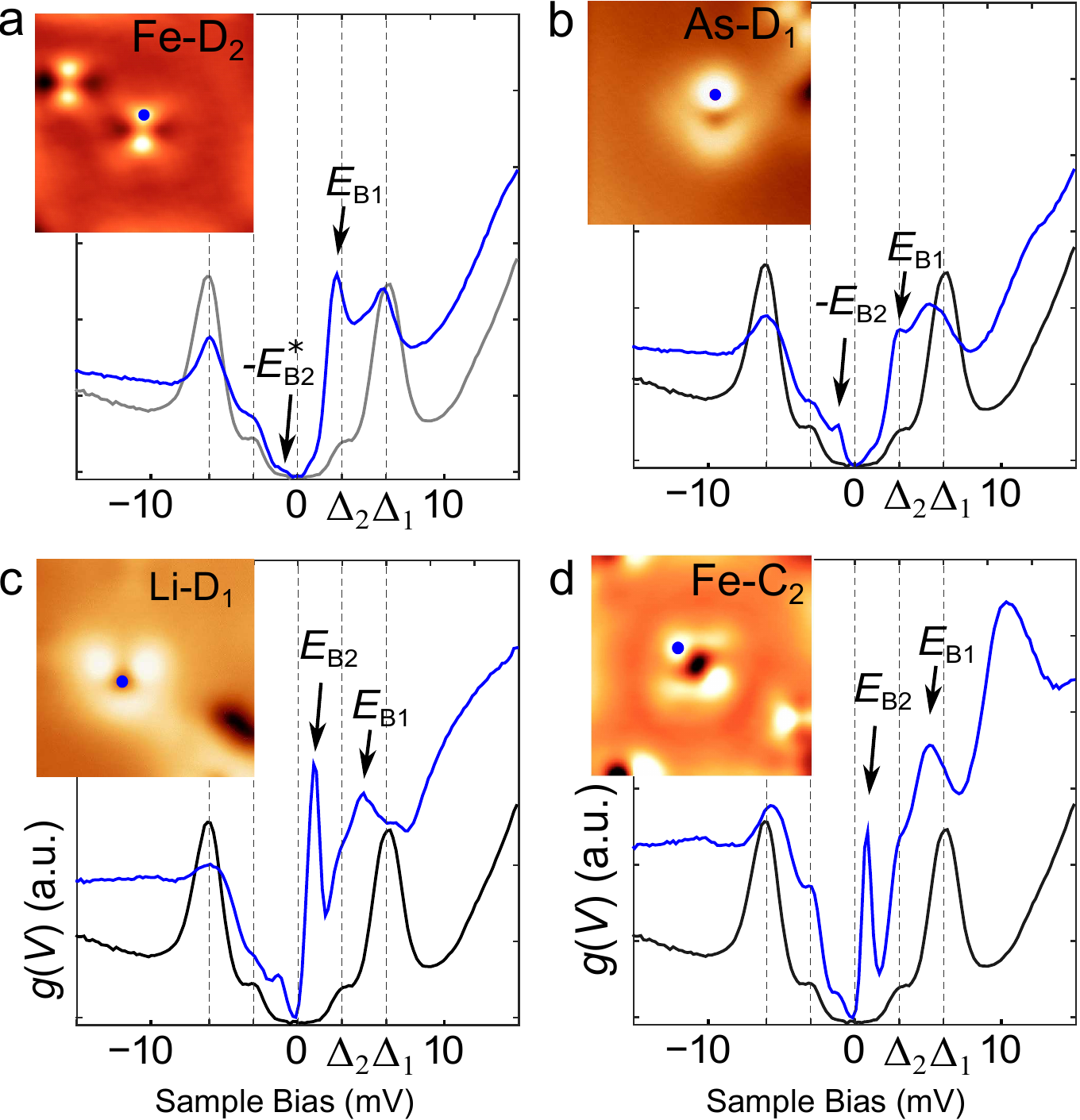}
\caption{\label{figS9_ExpSTS} 
\textbf{Tunneling spectra $g(V)$ on (a) Fe-D$_2$, (b) As-D$_1$, (c) Li-D$_{1}$, and (d) Fe-C$_2$ defects.} The black curve is the reference spectrum taken on a defect-free area. The locations for taking the defect spectra are marked as blue dots in the inserts. All data in this figure were taken in a home-built low temperature STM at 1.6~K\cite{white_stiff_2011}.
}
\end{figure}

Tunneling spectra obtained on four native defects are shown in Figure~\ref{figS9_ExpSTS}. The reference spectrum measured on a defect-free spot shows two superconducting gaps, $\Delta_1$ = 6~meV and $\Delta_{2}$ = 3~meV, consistent with previous results~\cite{ChiPRL2012,Allan2012,Hanaguri2012}. As shown in Figure~\ref{figS9_ExpSTS}(a), the Fe-D$_2$ defect produces a strong bound state corresponding to the in-gap bound state for the large gap. QPI of this bound state is used to determine the order parameter in LiFeAs (see Fig.~3 of Ref.~\onlinecite{ChiPRL2017}). The shallow shoulder feature at $E^{*}_{\mrm{B2}} \sim 1.3$~meV possibly corresponds to bound states of the small gaps. The other three types of native defects generate strong in-gap bound states, as seen in Figure~\ref{figS9_ExpSTS}(b-d)~\cite{GrothePRB}. These strong bound states, labeled as $E_{\mrm{B2}}$, give rise to the strongest signal in DBS-QPI at 1.2~meV and are associated with the bands at the small gap.

\section{Recovering phase information for scattering from multiple defects}
In the case of an area with many defects, the actual phase of the QPI for individual defects can be recovered~\cite{HAEM2017}. In the sparse case, a tunneling conductance map with multiple defects can be written as
\begin{equation}
g(\br,\omega) = \sum_{\mrm{\textbf{R}}_{i}} g_{S}(\br-\mrm{\textbf{R}}_{i}, \omega),
\label{eq:Multi-singleR}
\end{equation}
where $\mrm{\textbf{R}}_{i}$ is the location of the \textit{i}-th defect and $g_{S}(\br, \omega)$ is the tunneling conductance map for a single defect at the origin. Then the Fourier transformation is   
\begin{eqnarray}
\tilde g(\bq, \omega) &=& \int \mrm{d}\br\ e^{-\mrm{i}\bq\br} g(\br,\omega) \nonumber \\
 &=& \int \mrm{d}\br\ e^{-\mrm{i}\bq\br} \sum_{\mrm{\textbf{R}}_{i}} g_{S}(\br-\mrm{\textbf{R}}_{i}, \omega) \nonumber \\
 &=& \sum_{\mrm{\textbf{R}}_{i}} e^{\mrm{i}\bq\mrm{\textbf{R}}_{i}} \int \mrm{d}\br\ e^{-\mrm{i}\bq\br} g_{S}(\br, \omega) \nonumber \\
 &=& \sum_{\mrm{\textbf{R}}_{i}} e^{\mrm{i}\bq\mrm{\textbf{R}}_{i}} \tilde g_{S}(\bq, \omega),
\label{eq:Multi-singleQ}
\end{eqnarray}
where $\tilde g_{S}(\bq, \omega)$ is the QPI of a single defect in $\bq$-space. The prefactor $\sum_{\mrm{\textbf{R}}_{i}}e^{\mrm{i}\bq\mrm{\textbf{R}}_{i}}$ causes interference effects between defects, reducing the signal strength. Its absolute value is generally proportional to $\sqrt{N}$ instead of $N$, where $N$ is the number of defects. From  Equation~\ref{eq:Multi-singleQ}, the single defect QPI, $\tilde g_{S}(\bq, \omega)$, can be extracted using a multiple-defect-configuration correction
\begin{equation}
\tilde g_{S}(\bq,\omega) = \frac{\tilde g(\bq, \omega)}{\sum_{\mrm{\textbf{R}}_{i}}e^{\mrm{i}\bq\mrm{\textbf{R}}_{i}}}.
\label{eq:g_Single}
\end{equation}
This method only applies to maps with primarily one type of defect, in which $\tilde g_{S}(\bq,\omega)$ of all defects are identical.

\begin{figure}[ht]
\centering
\includegraphics[width=0.99\linewidth]{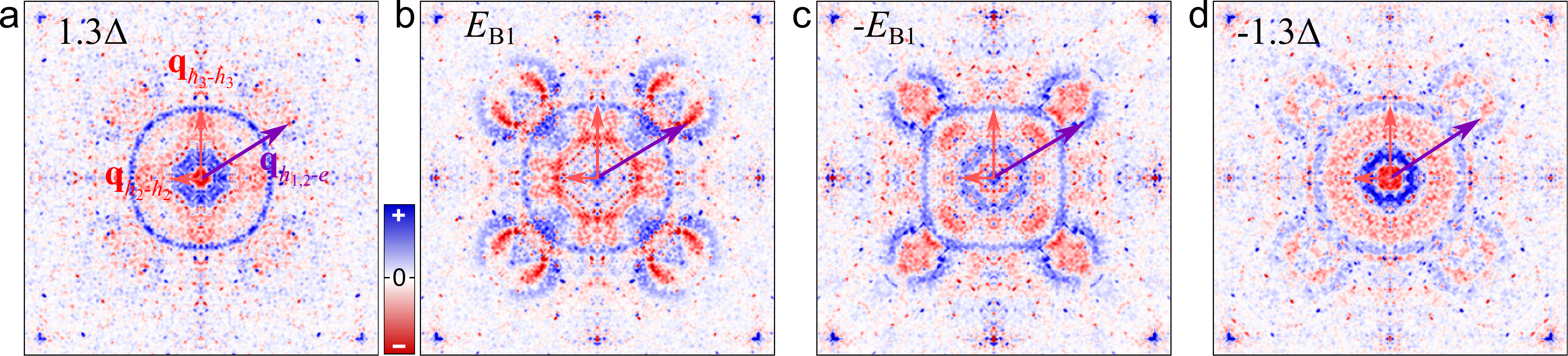}
\caption{\label{figS12ConCorrect} 
\textbf{QPI of all Fe-D$_2$ defects with multiple-defect-configuration correction.} (\textbf{a}) - (\textbf{d}) Re[$\tilde g_{S}(\bq,\omega)$] above the superconducting gap (1.3$\Delta_1$), at the bound state energies ($\pm E_{\mrm{B1}}$), and below the superconducting gap (-1.3$\Delta_1$).
}
\end{figure}

We extracted $\tilde g_{S}(\bq,\omega)$ for the Fe-D$_2$ defect by identifying the positions of all Fe-D$_2$ defects and masking the signals from the other defects. Because of the D$_2$ symmetry of the defect, signals exist only in the real part, Re[$\tilde g_{S}(\bq,\omega)$]. As shown in  Figure~\ref{figS12ConCorrect}, a map Re[$\tilde g_{S}(\bq,\omega)$] for a Fe-D$_2$ was reconstructed from two measured maps which contain multiple defects in the areas. The QPI features are consistent with the one obtained from a single defect (see  Figure~\ref{figS11_HAEM}b-c). However, multiple-defect-configuration correction gives enhanced $\bq$-space resolution. 
The $\bq_{h_{1,2}-e}$ QPI features is only enhanced at $\pm E_{\mrm{B}1}$ and has the sign-change between positive and negative bound state energies, in agreement with theoretical calculations using the $s_{\pm}$ order parameter. 

However, the experimental $\tilde g_{S}(\bq,\omega)$ has a more complicated phase than from the theoretical calculation using a $\delta$-potential, hence a direct comparison between experimental data and theoretical simulations is challenging. This is the reason that phase-referenced QPI and ratio maps were employed to analyze the scattering pattern here and in Ref.~\onlinecite{ChiPRL2017}. Phase-referenced QPI sets the zero phase for QPI at $E$ and highlights the phase changes at energy $-E$ with opposite polarity. Ratio-map QPI incorporates the phase contrast by taking the ratio between the positive and negative energies.

\bibliographystyle{apsrev4-1}
\label{Bibliography}
\bibliography{LiFeAs_BoundStateQPI}

\begin{thebibliography}{35}%
\makeatletter
\providecommand \@ifxundefined [1]{%
 \@ifx{#1\undefined}
}%
\providecommand \@ifnum [1]{%
 \ifnum #1\expandafter \@firstoftwo
 \else \expandafter \@secondoftwo
 \fi
}%
\providecommand \@ifx [1]{%
 \ifx #1\expandafter \@firstoftwo
 \else \expandafter \@secondoftwo
 \fi
}%
\providecommand \natexlab [1]{#1}%
\providecommand \enquote  [1]{``#1''}%
\providecommand \bibnamefont  [1]{#1}%
\providecommand \bibfnamefont [1]{#1}%
\providecommand \citenamefont [1]{#1}%
\providecommand \href@noop [0]{\@secondoftwo}%
\providecommand \href [0]{\begingroup \@sanitize@url \@href}%
\providecommand \@href[1]{\@@startlink{#1}\@@href}%
\providecommand \@@href[1]{\endgroup#1\@@endlink}%
\providecommand \@sanitize@url [0]{\catcode `\\12\catcode `\$12\catcode
  `\&12\catcode `\#12\catcode `\^12\catcode `\_12\catcode `\%12\relax}%
\providecommand \@@startlink[1]{}%
\providecommand \@@endlink[0]{}%
\providecommand \url  [0]{\begingroup\@sanitize@url \@url }%
\providecommand \@url [1]{\endgroup\@href {#1}{\urlprefix }}%
\providecommand \urlprefix  [0]{URL }%
\providecommand \Eprint [0]{\href }%
\providecommand \doibase [0]{http://dx.doi.org/}%
\providecommand \selectlanguage [0]{\@gobble}%
\providecommand \bibinfo  [0]{\@secondoftwo}%
\providecommand \bibfield  [0]{\@secondoftwo}%
\providecommand \translation [1]{[#1]}%
\providecommand \BibitemOpen [0]{}%
\providecommand \bibitemStop [0]{}%
\providecommand \bibitemNoStop [0]{.\EOS\space}%
\providecommand \EOS [0]{\spacefactor3000\relax}%
\providecommand \BibitemShut  [1]{\csname bibitem#1\endcsname}%
\let\auto@bib@innerbib\@empty
\bibitem [{\citenamefont {Hoffman}(2002)}]{hoffman_imaging_2002}%
  \BibitemOpen
  \bibfield  {author} {\bibinfo {author} {\bibfnamefont {J.~E.}\ \bibnamefont
  {Hoffman}},\ }\href {\doibase 10.1126/science.1072640} {\bibfield  {journal}
  {\bibinfo  {journal} {Science}\ }\textbf {\bibinfo {volume} {297}},\ \bibinfo
  {pages} {1148} (\bibinfo {year} {2002})}\BibitemShut {NoStop}%
\bibitem [{\citenamefont {Wang}\ and\ \citenamefont
  {Lee}(2003)}]{wang_quasiparticle_2003}%
  \BibitemOpen
  \bibfield  {author} {\bibinfo {author} {\bibfnamefont {Q.-H.}\ \bibnamefont
  {Wang}}\ and\ \bibinfo {author} {\bibfnamefont {D.-H.}\ \bibnamefont {Lee}},\
  }\href {\doibase 10.1103/PhysRevB.67.020511} {\bibfield  {journal} {\bibinfo
  {journal} {Phys. Rev. B}\ }\textbf {\bibinfo {volume} {67}},\ \bibinfo
  {pages} {020511} (\bibinfo {year} {2003})}\BibitemShut {NoStop}%
\bibitem [{\citenamefont {McElroy}\ \emph {et~al.}(2003)\citenamefont
  {McElroy}, \citenamefont {Simmonds}, \citenamefont {Hoffman}, \citenamefont
  {Lee}, \citenamefont {Orenstein}, \citenamefont {Eisaki}, \citenamefont
  {Uchida},\ and\ \citenamefont {Davis}}]{mcelroy_relating_2003}%
  \BibitemOpen
  \bibfield  {author} {\bibinfo {author} {\bibfnamefont {K.}~\bibnamefont
  {McElroy}}, \bibinfo {author} {\bibfnamefont {R.~W.}\ \bibnamefont
  {Simmonds}}, \bibinfo {author} {\bibfnamefont {J.~E.}\ \bibnamefont
  {Hoffman}}, \bibinfo {author} {\bibfnamefont {D.-H.}\ \bibnamefont {Lee}},
  \bibinfo {author} {\bibfnamefont {J.}~\bibnamefont {Orenstein}}, \bibinfo
  {author} {\bibfnamefont {H.}~\bibnamefont {Eisaki}}, \bibinfo {author}
  {\bibfnamefont {S.}~\bibnamefont {Uchida}}, \ and\ \bibinfo {author}
  {\bibfnamefont {J.~C.}\ \bibnamefont {Davis}},\ }\href {\doibase
  10.1038/nature01496} {\bibfield  {journal} {\bibinfo  {journal} {Nature}\
  }\textbf {\bibinfo {volume} {422}},\ \bibinfo {pages} {592} (\bibinfo {year}
  {2003})}\BibitemShut {NoStop}%
\bibitem [{\citenamefont {Hanaguri}\ \emph {et~al.}(2007)\citenamefont
  {Hanaguri}, \citenamefont {Kohsaka}, \citenamefont {Davis}, \citenamefont
  {Lupien}, \citenamefont {Yamada}, \citenamefont {Azuma}, \citenamefont
  {Takano}, \citenamefont {Ohishi}, \citenamefont {Ono},\ and\ \citenamefont
  {Takagi}}]{hanaguri_quasiparticle_2007}%
  \BibitemOpen
  \bibfield  {author} {\bibinfo {author} {\bibfnamefont {T.}~\bibnamefont
  {Hanaguri}}, \bibinfo {author} {\bibfnamefont {Y.}~\bibnamefont {Kohsaka}},
  \bibinfo {author} {\bibfnamefont {J.~C.}\ \bibnamefont {Davis}}, \bibinfo
  {author} {\bibfnamefont {C.}~\bibnamefont {Lupien}}, \bibinfo {author}
  {\bibfnamefont {I.}~\bibnamefont {Yamada}}, \bibinfo {author} {\bibfnamefont
  {M.}~\bibnamefont {Azuma}}, \bibinfo {author} {\bibfnamefont
  {M.}~\bibnamefont {Takano}}, \bibinfo {author} {\bibfnamefont
  {K.}~\bibnamefont {Ohishi}}, \bibinfo {author} {\bibfnamefont
  {M.}~\bibnamefont {Ono}}, \ and\ \bibinfo {author} {\bibfnamefont
  {H.}~\bibnamefont {Takagi}},\ }\href {\doibase 10.1038/nphys753} {\bibfield
  {journal} {\bibinfo  {journal} {Nature Physics}\ }\textbf {\bibinfo {volume}
  {3}},\ \bibinfo {pages} {865} (\bibinfo {year} {2007})}\BibitemShut {NoStop}%
\bibitem [{\citenamefont {Rutter}\ \emph {et~al.}(2007)\citenamefont {Rutter},
  \citenamefont {Crain}, \citenamefont {Guisinger}, \citenamefont {Li},
  \citenamefont {First},\ and\ \citenamefont {Stroscio}}]{Rutter2007}%
  \BibitemOpen
  \bibfield  {author} {\bibinfo {author} {\bibfnamefont {G.~M.}\ \bibnamefont
  {Rutter}}, \bibinfo {author} {\bibfnamefont {J.~N.}\ \bibnamefont {Crain}},
  \bibinfo {author} {\bibfnamefont {N.~P.}\ \bibnamefont {Guisinger}}, \bibinfo
  {author} {\bibfnamefont {T.}~\bibnamefont {Li}}, \bibinfo {author}
  {\bibfnamefont {P.~N.}\ \bibnamefont {First}}, \ and\ \bibinfo {author}
  {\bibfnamefont {J.~A.}\ \bibnamefont {Stroscio}},\ }\href {\doibase
  10.1126/science.1142882} {\bibfield  {journal} {\bibinfo  {journal}
  {Science}\ }\textbf {\bibinfo {volume} {317}},\ \bibinfo {pages} {219}
  (\bibinfo {year} {2007})}\BibitemShut {NoStop}%
\bibitem [{\citenamefont {Kohsaka}\ \emph {et~al.}(2008)\citenamefont
  {Kohsaka}, \citenamefont {Taylor}, \citenamefont {Wahl}, \citenamefont
  {Schmidt}, \citenamefont {Lee}, \citenamefont {Fujita}, \citenamefont
  {Alldredge}, \citenamefont {McElroy}, \citenamefont {Lee}, \citenamefont
  {Eisaki}, \citenamefont {Uchida}, \citenamefont {Lee},\ and\ \citenamefont
  {Davis}}]{kohsaka_how_2008}%
  \BibitemOpen
  \bibfield  {author} {\bibinfo {author} {\bibfnamefont {Y.}~\bibnamefont
  {Kohsaka}}, \bibinfo {author} {\bibfnamefont {C.}~\bibnamefont {Taylor}},
  \bibinfo {author} {\bibfnamefont {P.}~\bibnamefont {Wahl}}, \bibinfo {author}
  {\bibfnamefont {A.}~\bibnamefont {Schmidt}}, \bibinfo {author} {\bibfnamefont
  {J.}~\bibnamefont {Lee}}, \bibinfo {author} {\bibfnamefont {K.}~\bibnamefont
  {Fujita}}, \bibinfo {author} {\bibfnamefont {J.~W.}\ \bibnamefont
  {Alldredge}}, \bibinfo {author} {\bibfnamefont {K.}~\bibnamefont {McElroy}},
  \bibinfo {author} {\bibfnamefont {J.}~\bibnamefont {Lee}}, \bibinfo {author}
  {\bibfnamefont {H.}~\bibnamefont {Eisaki}}, \bibinfo {author} {\bibfnamefont
  {S.}~\bibnamefont {Uchida}}, \bibinfo {author} {\bibfnamefont {D.-H.}\
  \bibnamefont {Lee}}, \ and\ \bibinfo {author} {\bibfnamefont {J.~C.}\
  \bibnamefont {Davis}},\ }\href {\doibase 10.1038/nature07243} {\bibfield
  {journal} {\bibinfo  {journal} {Nature}\ }\textbf {\bibinfo {volume} {454}},\
  \bibinfo {pages} {1072} (\bibinfo {year} {2008})}\BibitemShut {NoStop}%
\bibitem [{\citenamefont {Roushan}\ \emph {et~al.}(2009)\citenamefont
  {Roushan}, \citenamefont {Seo}, \citenamefont {Parker}, \citenamefont {Hor},
  \citenamefont {Hsieh}, \citenamefont {Qian}, \citenamefont {Richardella},
  \citenamefont {Hasan}, \citenamefont {Cava},\ and\ \citenamefont
  {Yazdani}}]{Roushan2009}%
  \BibitemOpen
  \bibfield  {author} {\bibinfo {author} {\bibfnamefont {P.}~\bibnamefont
  {Roushan}}, \bibinfo {author} {\bibfnamefont {J.}~\bibnamefont {Seo}},
  \bibinfo {author} {\bibfnamefont {C.~V.}\ \bibnamefont {Parker}}, \bibinfo
  {author} {\bibfnamefont {Y.~S.}\ \bibnamefont {Hor}}, \bibinfo {author}
  {\bibfnamefont {D.}~\bibnamefont {Hsieh}}, \bibinfo {author} {\bibfnamefont
  {D.}~\bibnamefont {Qian}}, \bibinfo {author} {\bibfnamefont {A.}~\bibnamefont
  {Richardella}}, \bibinfo {author} {\bibfnamefont {M.~Z.}\ \bibnamefont
  {Hasan}}, \bibinfo {author} {\bibfnamefont {R.~J.}\ \bibnamefont {Cava}}, \
  and\ \bibinfo {author} {\bibfnamefont {A.}~\bibnamefont {Yazdani}},\ }\href
  {\doibase 10.1038/nature08308} {\bibfield  {journal} {\bibinfo  {journal}
  {Nature}\ }\textbf {\bibinfo {volume} {460}},\ \bibinfo {pages} {1106}
  (\bibinfo {year} {2009})}\BibitemShut {NoStop}%
\bibitem [{\citenamefont {Hanaguri}\ \emph {et~al.}(2009)\citenamefont
  {Hanaguri}, \citenamefont {Kohsaka}, \citenamefont {Ono}, \citenamefont
  {Maltseva}, \citenamefont {Coleman}, \citenamefont {Yamada}, \citenamefont
  {Azuma}, \citenamefont {Takano}, \citenamefont {Ohishi},\ and\ \citenamefont
  {Takagi}}]{Hanaguri2009Dwave}%
  \BibitemOpen
  \bibfield  {author} {\bibinfo {author} {\bibfnamefont {T.}~\bibnamefont
  {Hanaguri}}, \bibinfo {author} {\bibfnamefont {Y.}~\bibnamefont {Kohsaka}},
  \bibinfo {author} {\bibfnamefont {M.}~\bibnamefont {Ono}}, \bibinfo {author}
  {\bibfnamefont {M.}~\bibnamefont {Maltseva}}, \bibinfo {author}
  {\bibfnamefont {P.}~\bibnamefont {Coleman}}, \bibinfo {author} {\bibfnamefont
  {I.}~\bibnamefont {Yamada}}, \bibinfo {author} {\bibfnamefont
  {M.}~\bibnamefont {Azuma}}, \bibinfo {author} {\bibfnamefont
  {M.}~\bibnamefont {Takano}}, \bibinfo {author} {\bibfnamefont
  {K.}~\bibnamefont {Ohishi}}, \ and\ \bibinfo {author} {\bibfnamefont
  {H.}~\bibnamefont {Takagi}},\ }\href {\doibase 10.1126/science.1166138}
  {\bibfield  {journal} {\bibinfo  {journal} {Science}\ }\textbf {\bibinfo
  {volume} {323}},\ \bibinfo {pages} {923} (\bibinfo {year}
  {2009})}\BibitemShut {NoStop}%
\bibitem [{\citenamefont {Hanaguri}\ \emph
  {et~al.}(2010{\natexlab{a}})\citenamefont {Hanaguri}, \citenamefont
  {Niitaka}, \citenamefont {Kuroki},\ and\ \citenamefont
  {Takagi}}]{Hanaguri2010}%
  \BibitemOpen
  \bibfield  {author} {\bibinfo {author} {\bibfnamefont {T.}~\bibnamefont
  {Hanaguri}}, \bibinfo {author} {\bibfnamefont {S.}~\bibnamefont {Niitaka}},
  \bibinfo {author} {\bibfnamefont {K.}~\bibnamefont {Kuroki}}, \ and\ \bibinfo
  {author} {\bibfnamefont {H.}~\bibnamefont {Takagi}},\ }\href {\doibase
  10.1126/science.1187399} {\bibfield  {journal} {\bibinfo  {journal}
  {Science}\ }\textbf {\bibinfo {volume} {328}},\ \bibinfo {pages} {474}
  (\bibinfo {year} {2010}{\natexlab{a}})}\BibitemShut {NoStop}%
\bibitem [{\citenamefont {{T. H\"anke, S. Sykora, R. Schlegel, D. Baumann, L.
  Harnagea, S. Wurmehl, M. Daghofer, B. B\"uchner, J. van den Brink, and C.
  Hess}}(2012)}]{Hess-STM}%
  \BibitemOpen
  \bibfield  {author} {\bibinfo {author} {\bibnamefont {{T. H\"anke, S. Sykora,
  R. Schlegel, D. Baumann, L. Harnagea, S. Wurmehl, M. Daghofer, B. B\"uchner,
  J. van den Brink, and C. Hess}}},\ }\href {\doibase
  10.1103/PhysRevLett.108.127001} {\bibfield  {journal} {\bibinfo  {journal}
  {Phys. Rev. Lett.}\ }\textbf {\bibinfo {volume} {108}},\ \bibinfo {pages}
  {127001} (\bibinfo {year} {2012})}\BibitemShut {NoStop}%
\bibitem [{\citenamefont {{M. P. Allan, A. W. Rost, A. P. Mackenzie, Yang Xie,
  J. C. Davis, K. Kihou, C. H. Lee, A. Iyo, H. Eisaki, T.-M.
  Chuang}}(2012)}]{Allan2012}%
  \BibitemOpen
  \bibfield  {author} {\bibinfo {author} {\bibnamefont {{M. P. Allan, A. W.
  Rost, A. P. Mackenzie, Yang Xie, J. C. Davis, K. Kihou, C. H. Lee, A. Iyo, H.
  Eisaki, T.-M. Chuang}}},\ }\href {\doibase 10.1126/science.1218726}
  {\bibfield  {journal} {\bibinfo  {journal} {Science}\ }\textbf {\bibinfo
  {volume} {336}},\ \bibinfo {pages} {563} (\bibinfo {year}
  {2012})}\BibitemShut {NoStop}%
\bibitem [{\citenamefont {Allan}\ \emph {et~al.}(2013)\citenamefont {Allan},
  \citenamefont {Massee}, \citenamefont {Morr}, \citenamefont {Van~Dyke},
  \citenamefont {Rost}, \citenamefont {Mackenzie}, \citenamefont {Petrovic},\
  and\ \citenamefont {Davis}}]{Allan2013CeCoIn5}%
  \BibitemOpen
  \bibfield  {author} {\bibinfo {author} {\bibfnamefont {M.~P.}\ \bibnamefont
  {Allan}}, \bibinfo {author} {\bibfnamefont {F.}~\bibnamefont {Massee}},
  \bibinfo {author} {\bibfnamefont {D.~K.}\ \bibnamefont {Morr}}, \bibinfo
  {author} {\bibfnamefont {J.}~\bibnamefont {Van~Dyke}}, \bibinfo {author}
  {\bibfnamefont {A.~W.}\ \bibnamefont {Rost}}, \bibinfo {author}
  {\bibfnamefont {A.~P.}\ \bibnamefont {Mackenzie}}, \bibinfo {author}
  {\bibfnamefont {C.}~\bibnamefont {Petrovic}}, \ and\ \bibinfo {author}
  {\bibfnamefont {J.~C.}\ \bibnamefont {Davis}},\ }\href
  {http://dx.doi.org/10.1038/nphys2671} {\bibfield  {journal} {\bibinfo
  {journal} {Nature Physics}\ }\textbf {\bibinfo {volume} {9}},\ \bibinfo
  {pages} {468} (\bibinfo {year} {2013})}\BibitemShut {NoStop}%
\bibitem [{\citenamefont {Zhou}\ \emph {et~al.}(2013)\citenamefont {Zhou},
  \citenamefont {Misra}, \citenamefont {da~Silva~Neto}, \citenamefont
  {Aynajian}, \citenamefont {Baumbach}, \citenamefont {Thompson}, \citenamefont
  {Bauer},\ and\ \citenamefont {Yazdani}}]{Zhou2013CeCoIn5}%
  \BibitemOpen
  \bibfield  {author} {\bibinfo {author} {\bibfnamefont {B.~B.}\ \bibnamefont
  {Zhou}}, \bibinfo {author} {\bibfnamefont {S.}~\bibnamefont {Misra}},
  \bibinfo {author} {\bibfnamefont {E.~H.}\ \bibnamefont {da~Silva~Neto}},
  \bibinfo {author} {\bibfnamefont {P.}~\bibnamefont {Aynajian}}, \bibinfo
  {author} {\bibfnamefont {R.~E.}\ \bibnamefont {Baumbach}}, \bibinfo {author}
  {\bibfnamefont {J.~D.}\ \bibnamefont {Thompson}}, \bibinfo {author}
  {\bibfnamefont {E.~D.}\ \bibnamefont {Bauer}}, \ and\ \bibinfo {author}
  {\bibfnamefont {A.}~\bibnamefont {Yazdani}},\ }\href
  {http://dx.doi.org/10.1038/nphys2672} {\bibfield  {journal} {\bibinfo
  {journal} {Nature Physics}\ }\textbf {\bibinfo {volume} {9}},\ \bibinfo
  {pages} {474} (\bibinfo {year} {2013})}\BibitemShut {NoStop}%
\bibitem [{\citenamefont {da~Silva~Neto}\ \emph {et~al.}(2014)\citenamefont
  {da~Silva~Neto}, \citenamefont {Aynajian}, \citenamefont {Frano},
  \citenamefont {Comin}, \citenamefont {Schierle}, \citenamefont {Weschke},
  \citenamefont {Gyenis}, \citenamefont {Wen}, \citenamefont {Schneeloch},
  \citenamefont {Xu}, \citenamefont {Ono}, \citenamefont {Gu}, \citenamefont
  {Le~Tacon},\ and\ \citenamefont {Yazdani}}]{Neto2013}%
  \BibitemOpen
  \bibfield  {author} {\bibinfo {author} {\bibfnamefont {E.~H.}\ \bibnamefont
  {da~Silva~Neto}}, \bibinfo {author} {\bibfnamefont {P.}~\bibnamefont
  {Aynajian}}, \bibinfo {author} {\bibfnamefont {A.}~\bibnamefont {Frano}},
  \bibinfo {author} {\bibfnamefont {R.}~\bibnamefont {Comin}}, \bibinfo
  {author} {\bibfnamefont {E.}~\bibnamefont {Schierle}}, \bibinfo {author}
  {\bibfnamefont {E.}~\bibnamefont {Weschke}}, \bibinfo {author} {\bibfnamefont
  {A.}~\bibnamefont {Gyenis}}, \bibinfo {author} {\bibfnamefont
  {J.}~\bibnamefont {Wen}}, \bibinfo {author} {\bibfnamefont {J.}~\bibnamefont
  {Schneeloch}}, \bibinfo {author} {\bibfnamefont {Z.}~\bibnamefont {Xu}},
  \bibinfo {author} {\bibfnamefont {S.}~\bibnamefont {Ono}}, \bibinfo {author}
  {\bibfnamefont {G.}~\bibnamefont {Gu}}, \bibinfo {author} {\bibfnamefont
  {M.}~\bibnamefont {Le~Tacon}}, \ and\ \bibinfo {author} {\bibfnamefont
  {A.}~\bibnamefont {Yazdani}},\ }\href {\doibase 10.1126/science.1243479}
  {\bibfield  {journal} {\bibinfo  {journal} {Science}\ }\textbf {\bibinfo
  {volume} {343}},\ \bibinfo {pages} {393} (\bibinfo {year}
  {2014})}\BibitemShut {NoStop}%
\bibitem [{\citenamefont {Rosenthal}\ \emph {et~al.}(2014)\citenamefont
  {Rosenthal}, \citenamefont {Andrade}, \citenamefont {Arguello}, \citenamefont
  {Fernandes}, \citenamefont {Xing}, \citenamefont {Wang}, \citenamefont {Jin},
  \citenamefont {Millis},\ and\ \citenamefont {Pasupathy}}]{Rosenthal2014}%
  \BibitemOpen
  \bibfield  {author} {\bibinfo {author} {\bibfnamefont {E.~P.}\ \bibnamefont
  {Rosenthal}}, \bibinfo {author} {\bibfnamefont {E.~F.}\ \bibnamefont
  {Andrade}}, \bibinfo {author} {\bibfnamefont {C.~J.}\ \bibnamefont
  {Arguello}}, \bibinfo {author} {\bibfnamefont {R.~M.}\ \bibnamefont
  {Fernandes}}, \bibinfo {author} {\bibfnamefont {L.~Y.}\ \bibnamefont {Xing}},
  \bibinfo {author} {\bibfnamefont {X.~C.}\ \bibnamefont {Wang}}, \bibinfo
  {author} {\bibfnamefont {C.~Q.}\ \bibnamefont {Jin}}, \bibinfo {author}
  {\bibfnamefont {A.~J.}\ \bibnamefont {Millis}}, \ and\ \bibinfo {author}
  {\bibfnamefont {A.~N.}\ \bibnamefont {Pasupathy}},\ }\href
  {http://dx.doi.org/10.1038/nphys2870} {\bibfield  {journal} {\bibinfo
  {journal} {Nature Physics}\ }\textbf {\bibinfo {volume} {10}},\ \bibinfo
  {pages} {225} (\bibinfo {year} {2014})}\BibitemShut {NoStop}%
\bibitem [{\citenamefont {Fan}\ \emph {et~al.}(2015)\citenamefont {Fan},
  \citenamefont {Zhang}, \citenamefont {Liu}, \citenamefont {Yan},
  \citenamefont {Ren}, \citenamefont {Peng}, \citenamefont {Xu}, \citenamefont
  {Xie}, \citenamefont {Hu}, \citenamefont {Zhang},\ and\ \citenamefont
  {Feng}}]{Fan2015}%
  \BibitemOpen
  \bibfield  {author} {\bibinfo {author} {\bibfnamefont {Q.}~\bibnamefont
  {Fan}}, \bibinfo {author} {\bibfnamefont {W.~H.}\ \bibnamefont {Zhang}},
  \bibinfo {author} {\bibfnamefont {X.}~\bibnamefont {Liu}}, \bibinfo {author}
  {\bibfnamefont {Y.~J.}\ \bibnamefont {Yan}}, \bibinfo {author} {\bibfnamefont
  {M.~Q.}\ \bibnamefont {Ren}}, \bibinfo {author} {\bibfnamefont
  {R.}~\bibnamefont {Peng}}, \bibinfo {author} {\bibfnamefont {H.~C.}\
  \bibnamefont {Xu}}, \bibinfo {author} {\bibfnamefont {B.~P.}\ \bibnamefont
  {Xie}}, \bibinfo {author} {\bibfnamefont {J.~P.}\ \bibnamefont {Hu}},
  \bibinfo {author} {\bibfnamefont {T.}~\bibnamefont {Zhang}}, \ and\ \bibinfo
  {author} {\bibfnamefont {D.~L.}\ \bibnamefont {Feng}},\ }\href {\doibase
  10.1038/nphys3450} {\bibfield  {journal} {\bibinfo  {journal} {Nature
  Physics}\ }\textbf {\bibinfo {volume} {11}},\ \bibinfo {pages} {946}
  (\bibinfo {year} {2015})}\BibitemShut {NoStop}%
\bibitem [{\citenamefont {Inoue}\ \emph {et~al.}(2016)\citenamefont {Inoue},
  \citenamefont {Gyenis}, \citenamefont {Wang}, \citenamefont {Li},
  \citenamefont {Oh}, \citenamefont {Jiang}, \citenamefont {Ni}, \citenamefont
  {Bernevig},\ and\ \citenamefont {Yazdani}}]{Inoue2016}%
  \BibitemOpen
  \bibfield  {author} {\bibinfo {author} {\bibfnamefont {H.}~\bibnamefont
  {Inoue}}, \bibinfo {author} {\bibfnamefont {A.}~\bibnamefont {Gyenis}},
  \bibinfo {author} {\bibfnamefont {Z.}~\bibnamefont {Wang}}, \bibinfo {author}
  {\bibfnamefont {J.}~\bibnamefont {Li}}, \bibinfo {author} {\bibfnamefont
  {S.~W.}\ \bibnamefont {Oh}}, \bibinfo {author} {\bibfnamefont
  {S.}~\bibnamefont {Jiang}}, \bibinfo {author} {\bibfnamefont
  {N.}~\bibnamefont {Ni}}, \bibinfo {author} {\bibfnamefont {B.~A.}\
  \bibnamefont {Bernevig}}, \ and\ \bibinfo {author} {\bibfnamefont
  {A.}~\bibnamefont {Yazdani}},\ }\href {\doibase 10.1126/science.aad8766}
  {\bibfield  {journal} {\bibinfo  {journal} {Science}\ }\textbf {\bibinfo
  {volume} {351}},\ \bibinfo {pages} {1184} (\bibinfo {year}
  {2016})}\BibitemShut {NoStop}%
\bibitem [{\citenamefont {Mazin}\ and\ \citenamefont
  {Singh}(2010)}]{Hanaguri2010comment}%
  \BibitemOpen
  \bibfield  {author} {\bibinfo {author} {\bibfnamefont {I.}~\bibnamefont
  {Mazin}}\ and\ \bibinfo {author} {\bibfnamefont {D.}~\bibnamefont {Singh}},\
  }\href@noop {} {\bibfield  {journal} {\bibinfo  {journal} {arXiv preprint
  arXiv:1007.0047}\ } (\bibinfo {year} {2010})}\BibitemShut {NoStop}%
\bibitem [{\citenamefont {Hanaguri}\ \emph
  {et~al.}(2010{\natexlab{b}})\citenamefont {Hanaguri}, \citenamefont
  {Niitaka}, \citenamefont {Kuroki},\ and\ \citenamefont
  {Takagi}}]{Hanaguri2010commentReply}%
  \BibitemOpen
  \bibfield  {author} {\bibinfo {author} {\bibfnamefont {T.}~\bibnamefont
  {Hanaguri}}, \bibinfo {author} {\bibfnamefont {S.}~\bibnamefont {Niitaka}},
  \bibinfo {author} {\bibfnamefont {K.}~\bibnamefont {Kuroki}}, \ and\ \bibinfo
  {author} {\bibfnamefont {H.}~\bibnamefont {Takagi}},\ }\href@noop {}
  {\bibfield  {journal} {\bibinfo  {journal} {arXiv preprint arXiv:1007.0307}\
  } (\bibinfo {year} {2010}{\natexlab{b}})}\BibitemShut {NoStop}%
\bibitem [{\citenamefont {Hirschfeld}\ \emph {et~al.}(2015)\citenamefont
  {Hirschfeld}, \citenamefont {Altenfeld}, \citenamefont {Eremin},\ and\
  \citenamefont {Mazin}}]{HAEM2015}%
  \BibitemOpen
  \bibfield  {author} {\bibinfo {author} {\bibfnamefont {P.~J.}\ \bibnamefont
  {Hirschfeld}}, \bibinfo {author} {\bibfnamefont {D.}~\bibnamefont
  {Altenfeld}}, \bibinfo {author} {\bibfnamefont {I.}~\bibnamefont {Eremin}}, \
  and\ \bibinfo {author} {\bibfnamefont {I.~I.}\ \bibnamefont {Mazin}},\ }\href
  {\doibase 10.1103/PhysRevB.92.184513} {\bibfield  {journal} {\bibinfo
  {journal} {Phys. Rev. B}\ }\textbf {\bibinfo {volume} {92}},\ \bibinfo
  {pages} {184513} (\bibinfo {year} {2015})}\BibitemShut {NoStop}%
\bibitem [{\citenamefont {Ménard}\ \emph {et~al.}(2015)\citenamefont
  {Ménard}, \citenamefont {Guissart}, \citenamefont {Brun}, \citenamefont
  {Pons}, \citenamefont {Stolyarov}, \citenamefont {Debontridder},
  \citenamefont {Leclerc}, \citenamefont {Janod}, \citenamefont {Cario},
  \citenamefont {Roditchev}, \citenamefont {Simon},\ and\ \citenamefont
  {Cren}}]{menard_coherent_2015}%
  \BibitemOpen
  \bibfield  {author} {\bibinfo {author} {\bibfnamefont {G.~C.}\ \bibnamefont
  {Ménard}}, \bibinfo {author} {\bibfnamefont {S.}~\bibnamefont {Guissart}},
  \bibinfo {author} {\bibfnamefont {C.}~\bibnamefont {Brun}}, \bibinfo {author}
  {\bibfnamefont {S.}~\bibnamefont {Pons}}, \bibinfo {author} {\bibfnamefont
  {V.~S.}\ \bibnamefont {Stolyarov}}, \bibinfo {author} {\bibfnamefont
  {F.}~\bibnamefont {Debontridder}}, \bibinfo {author} {\bibfnamefont {M.~V.}\
  \bibnamefont {Leclerc}}, \bibinfo {author} {\bibfnamefont {E.}~\bibnamefont
  {Janod}}, \bibinfo {author} {\bibfnamefont {L.}~\bibnamefont {Cario}},
  \bibinfo {author} {\bibfnamefont {D.}~\bibnamefont {Roditchev}}, \bibinfo
  {author} {\bibfnamefont {P.}~\bibnamefont {Simon}}, \ and\ \bibinfo {author}
  {\bibfnamefont {T.}~\bibnamefont {Cren}},\ }\href {\doibase
  10.1038/nphys3508} {\bibfield  {journal} {\bibinfo  {journal} {Nature
  Physics}\ }\textbf {\bibinfo {volume} {11}},\ \bibinfo {pages} {1013}
  (\bibinfo {year} {2015})}\BibitemShut {NoStop}%
\bibitem [{\citenamefont {Chi}\ \emph {et~al.}(2017)\citenamefont {Chi},
  \citenamefont {Hardy}, \citenamefont {Liang}, \citenamefont {Dosanjh},
  \citenamefont {Wahl}, \citenamefont {Burke},\ and\ \citenamefont
  {Bonn}}]{ChiPRL2017}%
  \BibitemOpen
  \bibfield  {author} {\bibinfo {author} {\bibfnamefont {S.}~\bibnamefont
  {Chi}}, \bibinfo {author} {\bibfnamefont {W.}~\bibnamefont {Hardy}}, \bibinfo
  {author} {\bibfnamefont {R.}~\bibnamefont {Liang}}, \bibinfo {author}
  {\bibfnamefont {P.}~\bibnamefont {Dosanjh}}, \bibinfo {author} {\bibfnamefont
  {P.}~\bibnamefont {Wahl}}, \bibinfo {author} {\bibfnamefont {S.}~\bibnamefont
  {Burke}}, \ and\ \bibinfo {author} {\bibfnamefont {D.}~\bibnamefont {Bonn}},\
  }\href@noop {} {\bibfield  {journal} {\bibinfo  {journal} {submitted}\ }
  (\bibinfo {year} {2017})}\BibitemShut {NoStop}%
\bibitem [{\citenamefont {Gastiasoro}\ \emph {et~al.}(2013)\citenamefont
  {Gastiasoro}, \citenamefont {Hirschfeld},\ and\ \citenamefont
  {Andersen}}]{GastiasoroPRB2013}%
  \BibitemOpen
  \bibfield  {author} {\bibinfo {author} {\bibfnamefont {M.~N.}\ \bibnamefont
  {Gastiasoro}}, \bibinfo {author} {\bibfnamefont {P.~J.}\ \bibnamefont
  {Hirschfeld}}, \ and\ \bibinfo {author} {\bibfnamefont {B.~M.}\ \bibnamefont
  {Andersen}},\ }\href {\doibase 10.1103/PhysRevB.88.220509} {\bibfield
  {journal} {\bibinfo  {journal} {Phys. Rev. B}\ }\textbf {\bibinfo {volume}
  {88}},\ \bibinfo {pages} {220509} (\bibinfo {year} {2013})}\BibitemShut
  {NoStop}%
\bibitem [{\citenamefont {{A. V. Balatsky, I. Vekhter and J.-X.
  Zhu}}(2006)}]{Balatsky2006review}%
  \BibitemOpen
  \bibfield  {author} {\bibinfo {author} {\bibnamefont {{A. V. Balatsky, I.
  Vekhter and J.-X. Zhu}}},\ }\href {\doibase 10.1103/RevModPhys.78.373}
  {\bibfield  {journal} {\bibinfo  {journal} {Rev. Mod. Phys.}\ }\textbf
  {\bibinfo {volume} {78}},\ \bibinfo {pages} {373} (\bibinfo {year}
  {2006})}\BibitemShut {NoStop}%
\bibitem [{\citenamefont {Tersoff}\ and\ \citenamefont
  {Hamann}(1983)}]{STMTheory_Tersoff}%
  \BibitemOpen
  \bibfield  {author} {\bibinfo {author} {\bibfnamefont {J.}~\bibnamefont
  {Tersoff}}\ and\ \bibinfo {author} {\bibfnamefont {D.~R.}\ \bibnamefont
  {Hamann}},\ }\href {\doibase 10.1103/PhysRevLett.50.1998} {\bibfield
  {journal} {\bibinfo  {journal} {Phys. Rev. Lett.}\ }\textbf {\bibinfo
  {volume} {50}},\ \bibinfo {pages} {1998} (\bibinfo {year}
  {1983})}\BibitemShut {NoStop}%
\bibitem [{\citenamefont {Chi}\ \emph {et~al.}(2016)\citenamefont {Chi},
  \citenamefont {Aluru}, \citenamefont {Singh}, \citenamefont {Liang},
  \citenamefont {Hardy}, \citenamefont {Bonn}, \citenamefont {Kreisel},
  \citenamefont {Andersen}, \citenamefont {Nelson}, \citenamefont {Berlijn},
  \citenamefont {Ku}, \citenamefont {Hirschfeld},\ and\ \citenamefont
  {Wahl}}]{ChiPRB2016}%
  \BibitemOpen
  \bibfield  {author} {\bibinfo {author} {\bibfnamefont {S.}~\bibnamefont
  {Chi}}, \bibinfo {author} {\bibfnamefont {R.}~\bibnamefont {Aluru}}, \bibinfo
  {author} {\bibfnamefont {U.~R.}\ \bibnamefont {Singh}}, \bibinfo {author}
  {\bibfnamefont {R.}~\bibnamefont {Liang}}, \bibinfo {author} {\bibfnamefont
  {W.~N.}\ \bibnamefont {Hardy}}, \bibinfo {author} {\bibfnamefont {D.~A.}\
  \bibnamefont {Bonn}}, \bibinfo {author} {\bibfnamefont {A.}~\bibnamefont
  {Kreisel}}, \bibinfo {author} {\bibfnamefont {B.~M.}\ \bibnamefont
  {Andersen}}, \bibinfo {author} {\bibfnamefont {R.}~\bibnamefont {Nelson}},
  \bibinfo {author} {\bibfnamefont {T.}~\bibnamefont {Berlijn}}, \bibinfo
  {author} {\bibfnamefont {W.}~\bibnamefont {Ku}}, \bibinfo {author}
  {\bibfnamefont {P.~J.}\ \bibnamefont {Hirschfeld}}, \ and\ \bibinfo {author}
  {\bibfnamefont {P.}~\bibnamefont {Wahl}},\ }\href {\doibase
  10.1103/PhysRevB.94.134515} {\bibfield  {journal} {\bibinfo  {journal} {Phys.
  Rev. B}\ }\textbf {\bibinfo {volume} {94}},\ \bibinfo {pages} {134515}
  (\bibinfo {year} {2016})}\BibitemShut {NoStop}%
\bibitem [{\citenamefont {Umezawa}\ \emph {et~al.}(2012)\citenamefont
  {Umezawa}, \citenamefont {Li}, \citenamefont {Miao}, \citenamefont
  {Nakayama}, \citenamefont {Liu}, \citenamefont {Richard}, \citenamefont
  {Sato}, \citenamefont {He}, \citenamefont {Wang}, \citenamefont {Chen},
  \citenamefont {Ding}, \citenamefont {Takahashi},\ and\ \citenamefont
  {Wang}}]{ARPES_Umezawa}%
  \BibitemOpen
  \bibfield  {author} {\bibinfo {author} {\bibfnamefont {K.}~\bibnamefont
  {Umezawa}}, \bibinfo {author} {\bibfnamefont {Y.}~\bibnamefont {Li}},
  \bibinfo {author} {\bibfnamefont {H.}~\bibnamefont {Miao}}, \bibinfo {author}
  {\bibfnamefont {K.}~\bibnamefont {Nakayama}}, \bibinfo {author}
  {\bibfnamefont {Z.-H.}\ \bibnamefont {Liu}}, \bibinfo {author} {\bibfnamefont
  {P.}~\bibnamefont {Richard}}, \bibinfo {author} {\bibfnamefont
  {T.}~\bibnamefont {Sato}}, \bibinfo {author} {\bibfnamefont {J.~B.}\
  \bibnamefont {He}}, \bibinfo {author} {\bibfnamefont {D.-M.}\ \bibnamefont
  {Wang}}, \bibinfo {author} {\bibfnamefont {G.~F.}\ \bibnamefont {Chen}},
  \bibinfo {author} {\bibfnamefont {H.}~\bibnamefont {Ding}}, \bibinfo {author}
  {\bibfnamefont {T.}~\bibnamefont {Takahashi}}, \ and\ \bibinfo {author}
  {\bibfnamefont {S.-C.}\ \bibnamefont {Wang}},\ }\href {\doibase
  10.1103/PhysRevLett.108.037002} {\bibfield  {journal} {\bibinfo  {journal}
  {Phys. Rev. Lett.}\ }\textbf {\bibinfo {volume} {108}},\ \bibinfo {pages}
  {037002} (\bibinfo {year} {2012})}\BibitemShut {NoStop}%
\bibitem [{\citenamefont {{T. Hanaguri, K. Kitagawa, K. Matsubayashi, Y.
  Mazaki, Y. Uwatoko, and H. Takagi}}(2012)}]{Hanaguri2012}%
  \BibitemOpen
  \bibfield  {author} {\bibinfo {author} {\bibnamefont {{T. Hanaguri, K.
  Kitagawa, K. Matsubayashi, Y. Mazaki, Y. Uwatoko, and H. Takagi}}},\ }\href
  {\doibase 10.1103/PhysRevB.85.214505} {\bibfield  {journal} {\bibinfo
  {journal} {Phys. Rev. B}\ }\textbf {\bibinfo {volume} {85}},\ \bibinfo
  {pages} {214505} (\bibinfo {year} {2012})}\BibitemShut {NoStop}%
\bibitem [{\citenamefont {Kreisel}\ \emph {et~al.}(2016)\citenamefont
  {Kreisel}, \citenamefont {Nelson}, \citenamefont {Berlijn}, \citenamefont
  {Ku}, \citenamefont {Aluru}, \citenamefont {Chi}, \citenamefont {Zhou},
  \citenamefont {Singh}, \citenamefont {Wahl}, \citenamefont {Liang},
  \citenamefont {Hardy}, \citenamefont {Bonn}, \citenamefont {Hirschfeld},\
  and\ \citenamefont {Andersen}}]{KreiselPRB2016}%
  \BibitemOpen
  \bibfield  {author} {\bibinfo {author} {\bibfnamefont {A.}~\bibnamefont
  {Kreisel}}, \bibinfo {author} {\bibfnamefont {R.}~\bibnamefont {Nelson}},
  \bibinfo {author} {\bibfnamefont {T.}~\bibnamefont {Berlijn}}, \bibinfo
  {author} {\bibfnamefont {W.}~\bibnamefont {Ku}}, \bibinfo {author}
  {\bibfnamefont {R.}~\bibnamefont {Aluru}}, \bibinfo {author} {\bibfnamefont
  {S.}~\bibnamefont {Chi}}, \bibinfo {author} {\bibfnamefont {H.}~\bibnamefont
  {Zhou}}, \bibinfo {author} {\bibfnamefont {U.~R.}\ \bibnamefont {Singh}},
  \bibinfo {author} {\bibfnamefont {P.}~\bibnamefont {Wahl}}, \bibinfo {author}
  {\bibfnamefont {R.}~\bibnamefont {Liang}}, \bibinfo {author} {\bibfnamefont
  {W.~N.}\ \bibnamefont {Hardy}}, \bibinfo {author} {\bibfnamefont {D.~A.}\
  \bibnamefont {Bonn}}, \bibinfo {author} {\bibfnamefont {P.~J.}\ \bibnamefont
  {Hirschfeld}}, \ and\ \bibinfo {author} {\bibfnamefont {B.~M.}\ \bibnamefont
  {Andersen}},\ }\href {\doibase 10.1103/PhysRevB.94.224518} {\bibfield
  {journal} {\bibinfo  {journal} {Phys. Rev. B}\ }\textbf {\bibinfo {volume}
  {94}},\ \bibinfo {pages} {224518} (\bibinfo {year} {2016})}\BibitemShut
  {NoStop}%
\bibitem [{\citenamefont {Sprau}\ \emph {et~al.}(2017)\citenamefont {Sprau},
  \citenamefont {Kostin}, \citenamefont {Kreisel}, \citenamefont {B{\"o}hmer},
  \citenamefont {Taufour}, \citenamefont {Canfield}, \citenamefont {Mukherjee},
  \citenamefont {Hirschfeld}, \citenamefont {Andersen},\ and\ \citenamefont
  {Davis}}]{sprau2016FeSe}%
  \BibitemOpen
  \bibfield  {author} {\bibinfo {author} {\bibfnamefont {P.~O.}\ \bibnamefont
  {Sprau}}, \bibinfo {author} {\bibfnamefont {A.}~\bibnamefont {Kostin}},
  \bibinfo {author} {\bibfnamefont {A.}~\bibnamefont {Kreisel}}, \bibinfo
  {author} {\bibfnamefont {A.~E.}\ \bibnamefont {B{\"o}hmer}}, \bibinfo
  {author} {\bibfnamefont {V.}~\bibnamefont {Taufour}}, \bibinfo {author}
  {\bibfnamefont {P.~C.}\ \bibnamefont {Canfield}}, \bibinfo {author}
  {\bibfnamefont {S.}~\bibnamefont {Mukherjee}}, \bibinfo {author}
  {\bibfnamefont {P.~J.}\ \bibnamefont {Hirschfeld}}, \bibinfo {author}
  {\bibfnamefont {B.~M.}\ \bibnamefont {Andersen}}, \ and\ \bibinfo {author}
  {\bibfnamefont {J.~C.~S.}\ \bibnamefont {Davis}},\ }\href {\doibase
  10.1126/science.aal1575} {\bibfield  {journal} {\bibinfo  {journal}
  {Science}\ }\textbf {\bibinfo {volume} {357}},\ \bibinfo {pages} {75}
  (\bibinfo {year} {2017})}\BibitemShut {NoStop}%
\bibitem [{\citenamefont {Martiny}\ \emph {et~al.}(2017)\citenamefont
  {Martiny}, \citenamefont {Kreisel}, \citenamefont {Hirschfeld},\ and\
  \citenamefont {Andersen}}]{HAEM2017}%
  \BibitemOpen
  \bibfield  {author} {\bibinfo {author} {\bibfnamefont {J.~H.~J.}\
  \bibnamefont {Martiny}}, \bibinfo {author} {\bibfnamefont {A.}~\bibnamefont
  {Kreisel}}, \bibinfo {author} {\bibfnamefont {P.~J.}\ \bibnamefont
  {Hirschfeld}}, \ and\ \bibinfo {author} {\bibfnamefont {B.~M.}\ \bibnamefont
  {Andersen}},\ }\href {\doibase 10.1103/PhysRevB.95.184507} {\bibfield
  {journal} {\bibinfo  {journal} {Phys. Rev. B}\ }\textbf {\bibinfo {volume}
  {95}},\ \bibinfo {pages} {184507} (\bibinfo {year} {2017})}\BibitemShut
  {NoStop}%
\bibitem [{\citenamefont {Raghu}\ \emph {et~al.}(2008)\citenamefont {Raghu},
  \citenamefont {Qi}, \citenamefont {Liu}, \citenamefont {Scalapino},\ and\
  \citenamefont {Zhang}}]{RaghuPRB2009_TwoOrb}%
  \BibitemOpen
  \bibfield  {author} {\bibinfo {author} {\bibfnamefont {S.}~\bibnamefont
  {Raghu}}, \bibinfo {author} {\bibfnamefont {X.-L.}\ \bibnamefont {Qi}},
  \bibinfo {author} {\bibfnamefont {C.-X.}\ \bibnamefont {Liu}}, \bibinfo
  {author} {\bibfnamefont {D.~J.}\ \bibnamefont {Scalapino}}, \ and\ \bibinfo
  {author} {\bibfnamefont {S.-C.}\ \bibnamefont {Zhang}},\ }\href {\doibase
  10.1103/PhysRevB.77.220503} {\bibfield  {journal} {\bibinfo  {journal} {Phys.
  Rev. B}\ }\textbf {\bibinfo {volume} {77}},\ \bibinfo {pages} {220503}
  (\bibinfo {year} {2008})}\BibitemShut {NoStop}%
\bibitem [{\citenamefont {White}\ \emph {et~al.}(2011)\citenamefont {White},
  \citenamefont {Singh},\ and\ \citenamefont {Wahl}}]{white_stiff_2011}%
  \BibitemOpen
  \bibfield  {author} {\bibinfo {author} {\bibfnamefont {S.~C.}\ \bibnamefont
  {White}}, \bibinfo {author} {\bibfnamefont {U.~R.}\ \bibnamefont {Singh}}, \
  and\ \bibinfo {author} {\bibfnamefont {P.}~\bibnamefont {Wahl}},\ }\href
  {http://scitation.aip.org/content/aip/journal/rsi/82/11/10.1063/1.3663611}
  {\bibfield  {journal} {\bibinfo  {journal} {Rev. Sci. Instrum.}\ }\textbf
  {\bibinfo {volume} {82}},\ \bibinfo {pages} {113708} (\bibinfo {year}
  {2011})}\BibitemShut {NoStop}%
\bibitem [{\citenamefont {Chi}\ \emph {et~al.}(2012)\citenamefont {Chi},
  \citenamefont {Grothe}, \citenamefont {Liang}, \citenamefont {Dosanjh},
  \citenamefont {Hardy}, \citenamefont {Burke}, \citenamefont {Bonn},\ and\
  \citenamefont {Pennec}}]{ChiPRL2012}%
  \BibitemOpen
  \bibfield  {author} {\bibinfo {author} {\bibfnamefont {S.}~\bibnamefont
  {Chi}}, \bibinfo {author} {\bibfnamefont {S.}~\bibnamefont {Grothe}},
  \bibinfo {author} {\bibfnamefont {R.}~\bibnamefont {Liang}}, \bibinfo
  {author} {\bibfnamefont {P.}~\bibnamefont {Dosanjh}}, \bibinfo {author}
  {\bibfnamefont {W.~N.}\ \bibnamefont {Hardy}}, \bibinfo {author}
  {\bibfnamefont {S.~A.}\ \bibnamefont {Burke}}, \bibinfo {author}
  {\bibfnamefont {D.~A.}\ \bibnamefont {Bonn}}, \ and\ \bibinfo {author}
  {\bibfnamefont {Y.}~\bibnamefont {Pennec}},\ }\href {\doibase
  10.1103/PhysRevLett.109.087002} {\bibfield  {journal} {\bibinfo  {journal}
  {Phys. Rev. Lett.}\ }\textbf {\bibinfo {volume} {109}},\ \bibinfo {pages}
  {087002} (\bibinfo {year} {2012})}\BibitemShut {NoStop}%
\bibitem [{\citenamefont {Grothe}\ \emph {et~al.}(2012)\citenamefont {Grothe},
  \citenamefont {Chi}, \citenamefont {Dosanjh}, \citenamefont {Liang},
  \citenamefont {Hardy}, \citenamefont {Burke}, \citenamefont {Bonn},\ and\
  \citenamefont {Pennec}}]{GrothePRB}%
  \BibitemOpen
  \bibfield  {author} {\bibinfo {author} {\bibfnamefont {S.}~\bibnamefont
  {Grothe}}, \bibinfo {author} {\bibfnamefont {S.}~\bibnamefont {Chi}},
  \bibinfo {author} {\bibfnamefont {P.}~\bibnamefont {Dosanjh}}, \bibinfo
  {author} {\bibfnamefont {R.}~\bibnamefont {Liang}}, \bibinfo {author}
  {\bibfnamefont {W.~N.}\ \bibnamefont {Hardy}}, \bibinfo {author}
  {\bibfnamefont {S.~A.}\ \bibnamefont {Burke}}, \bibinfo {author}
  {\bibfnamefont {D.~A.}\ \bibnamefont {Bonn}}, \ and\ \bibinfo {author}
  {\bibfnamefont {Y.}~\bibnamefont {Pennec}},\ }\href {\doibase
  10.1103/PhysRevB.86.174503} {\bibfield  {journal} {\bibinfo  {journal} {Phys.
  Rev. B}\ }\textbf {\bibinfo {volume} {86}},\ \bibinfo {pages} {174503}
  (\bibinfo {year} {2012})}\BibitemShut {NoStop}%
\end{thebibliography}%
\end{document}